\documentclass{emulateapj}
\newcommand{\ha}{H$\alpha$}

\newcommand{\smy}{$\rm {\it h}_{100}^{-2}~ M_\odot ~ yr^{-1}$}
\newcommand{\ewmin}{10\AA}

\newcommand{\h}{$h^{-1}_{100}$}
\newcommand{\ca}{CL1040} 
\newcommand{\cb}{CL1054$-$12}
\newcommand{\cc}{CL1216}
\newcommand{\cj}{CL~J0023$+$0423B}

\newcommand{\edi}{EDisCS}

\newcommand{\rtwo}{$R_{200}$}

\newcommand{\sfrm}{$\rm \Sigma SFR/M_{cl}$}
\newcommand{\ewr}{EW}
\newcommand{\mcl}{M$\rm_{cl}$}
\shorttitle{\ha \ SFRs of $z \simeq 0.75$ EDisCS Galaxy Clusters}
\shortauthors{Finn et al.}

\begin{document}
\title{H$\alpha$-Derived Star-Formation Rates For Three $z \simeq 0.75$ 
\edi \ Galaxy Clusters\altaffilmark{1}}

\author{Rose A. Finn\altaffilmark{2,3}, Dennis Zaritsky\altaffilmark{4}, 
Donald W. McCarthy, Jr.\altaffilmark{4}, Bianca Poggianti\altaffilmark{5}, Gregory Rudnick\altaffilmark{6}, Claire Halliday\altaffilmark{7,8,9}, Bo Milvang-Jensen\altaffilmark{8}, Roser Pell\'o\altaffilmark{10}, Luc Simard\altaffilmark{11}}

\email{rfinn@astro.umass.edu; dzaritsky, mccarthy@as.arizona.edu}
\altaffiltext{1}{Based on observations with the MMT Observatory,
a joint venture of the Smithsonian Astrophysical Observatory and
the University of Arizona.}
\altaffiltext{2}{NSF Astronomy and Astrophysics Postdoctoral Fellow}
\altaffiltext{3}{Department of Astronomy, University of Massachusetts, Amherst, MA  01003; Current address:  Department of Physics, 515 Loudon Rd, Siena College, Loudonville, NY  12211}
\altaffiltext{4}{Steward Observatory, 933 N. Cherry Ave., University of Arizona, Tucson, AZ  85721}
\altaffiltext{5}{Osservatorio Astronomico, vicolo dell'Osservatorio 5, 35122 Padova}
\altaffiltext{6}{NOAO, 950 N. Cherry Ave, Tucson, AZ 85721}
\altaffiltext{7}{Max-Planck-Institut fuer Astrophysik,
Karl-Schwarzschild-Str. 1,
D-85741 Garching,
Germany}
\altaffiltext{8}{Max-Planck Institut f\"ur extraterrestrische Physik, Giessenbacstrasse, D-885748 Garching, Germany}
\altaffiltext{9}{
Current address:  Universitaetssternwarte,
Geismarlandstr. 11,
D-37083 Goettingen,
Germany}
\altaffiltext{10}{Laboratoire d'Astrophysique, UMR 5572, Observatoire Midi-Pyrenees, 14 Avenue E., Belin, 31400 Toulouse, France}
\altaffiltext{11}{Herzberg Institute of Astrophysics, National Research Council of Canada, Victoria, British Columbia, Canada}

\begin{abstract}
We present \ha-derived star-formation rates (SFRs) for three $z \simeq 0.75$ 
galaxy clusters.  Our $1\sigma$ flux limit corresponds 
to a star-formation rate of 0.10-0.24~\smy, and our minimum reliable 
\ha~+ [N~II] rest-frame equivalent width is $10$~\AA. 
We show that \ha \ narrowband imaging is an 
efficient method for measuring star formation in distant clusters.
In two out of three clusters, we find that the fraction of star-forming 
galaxies increases with projected distance from the cluster center.
We also find that the fraction of star-forming galaxies decreases with
increasing local galaxy surface density in the same two clusters.  
We compare the median rate of star formation 
among star-forming cluster galaxies to a small sample of star-forming 
field galaxies from the
literature and find that the median cluster SFRs are $\sim 50$\% 
less than the median field SFR.
We characterize cluster evolution in terms of the mass-normalized 
integrated cluster SFR and find that the $z \simeq 0.75$ clusters 
have more SFR per cluster mass on average
than the $z \le 0.4$ clusters from the literature. 
The interpretation of this result is complicated
by the dependence of the mass-normalized SFR on cluster 
mass and the lack of sufficient
overlap in the mass ranges covered by the low and high redshift samples.  
We find that the fraction and luminosities of the brightest 
starburst galaxies at $z \simeq 0.75$ are consistent with their being
progenitors of the post-starburst galaxies at $z \simeq 0.45$ 
if the post-starburst phase 
lasts several ($\sim 5$) times longer than the starburst phase.
\end{abstract}


\section{INTRODUCTION}

Although the correlation between various galaxy properties and large-scale (R $\sim$ 1 Mpc)
environment  is well established ({Lewis} {et~al.} 2002; {G{\' o}mez} {et~al.} 2003; {Kauffmann} {et~al.} 2004), the degree 
to which environment drives galaxy evolution is not. In particular, it is unclear whether the
differences among galaxies are primarily driven by local environmental 
effects, such as interactions
and merger events, which are themselves correlated with the large scale environment,
or by physical processes that are only important in highly over-dense regions,
such as ram-pressure stripping and galaxy harassment. This question has been notoriously
difficult to answer because the more noticeable changes in galaxy properties happen
over large redshift baselines and because at high redshifts our generally 
available diagnostics, such as galaxy colors,
are a rather blunt tool with which to unravel the competing, and perhaps complicitous,
physical processes. We have begun a study of the most time-sensitive diagnostic
of the evolution of a galaxy's stellar population, namely its 
current star formation rate (SFR), 
at the highest redshifts (z $\sim 0.8$) for which significant samples of 
galaxy clusters currently exist.

Astronomers derive SFRs from continuum emission 
at ultraviolet, far-infrared, and radio wavelengths and line
emission at optical, infrared and sub-millimeter wavelengths
({Kennicutt} 1998, and references therein).
In the local universe, \ha \ emission ($\lambda$6563\AA) 
is the conventional standard by which
to gauge star formation ({Kennicutt} 1998) because it directly measures 
the ionizing flux of young
massive stars, is intrinsically the strongest optical emission line,
and is less sensitive to extinction and metallicity than 
the [O~II]$\lambda$3727 line.
Currently, most SFR studies of $z > 0.4$ clusters rely 
on [O~II] as a star-formation
indicator because it is accessible
in the optical window out to $z < 1.5$, yet the ratio of 
[O~II] to \ha \ varies by a factor of 25 among
galaxies with an RMS of a factor of 2.5 ({Jansen} {et~al.} 2001).
SFRs measured from the \ha \ emission
line are directly comparable to $z < 0.4$ studies, and this is
essential in order to separate systematic from evolutionary effects.

To observe the \ha\ line, 
we are undertaking a near-infrared, narrowband \ha \ imaging
survey of ten $z \sim 0.8$ clusters.  We presented results for the first
cluster in our sample in {Finn} {et~al.} (2004), hereafter Paper I. 
Here we present results for three additional $z \simeq 0.75$ clusters from the 
ESO Distant Cluster Survey (\edi): CL~1040.7$-$1155 (\ca) at $z = 0.704$, 
CL~1054.7$-$1245 (\cb) at $z = 0.748$, and CL~1216.8$-$1201 (\cc) at $z = 0.794$.  
The \edi \ project is an ESO Large Programme studying 20 $0.4 < z < 0.8$ 
optically-selected clusters drawn from the 
Las Campanas Distant Cluster
Survey ({Gonzalez} {et~al.} 2001).  
The \edi \ collaboration has secured VRIJK imaging ({White} {et~al.} 2004) 
and spectroscopy for 30-66 members per cluster ({Halliday} {et~al.} 2004).
These ancillary data provide a powerful complement to our \ha \ imaging,
in terms of both calibration and interpretation of results.

This paper is organized as follows.
In \S\ref{comp_obs} we describe the observations and data reduction.  
In \S\ref{contsub} we describe how we measure the continuum in the
narrowband filter, and in 
\S\ref{results} we present results.  We compare our observations with
lower redshift cluster surveys in \S\ref{discussion}, and
we summarize in \S\ref{summary}.  
We assume $\Omega_0 = 0.3$ and $\Omega_\Lambda = 0.7$ throughout 
and express results in terms of $h_{100} = \rm \frac{H_0}{100~km/s/Mpc}$.

\section{Observations \& Data Reduction}
\label{comp_obs}
\begin{figure}
\plottwo{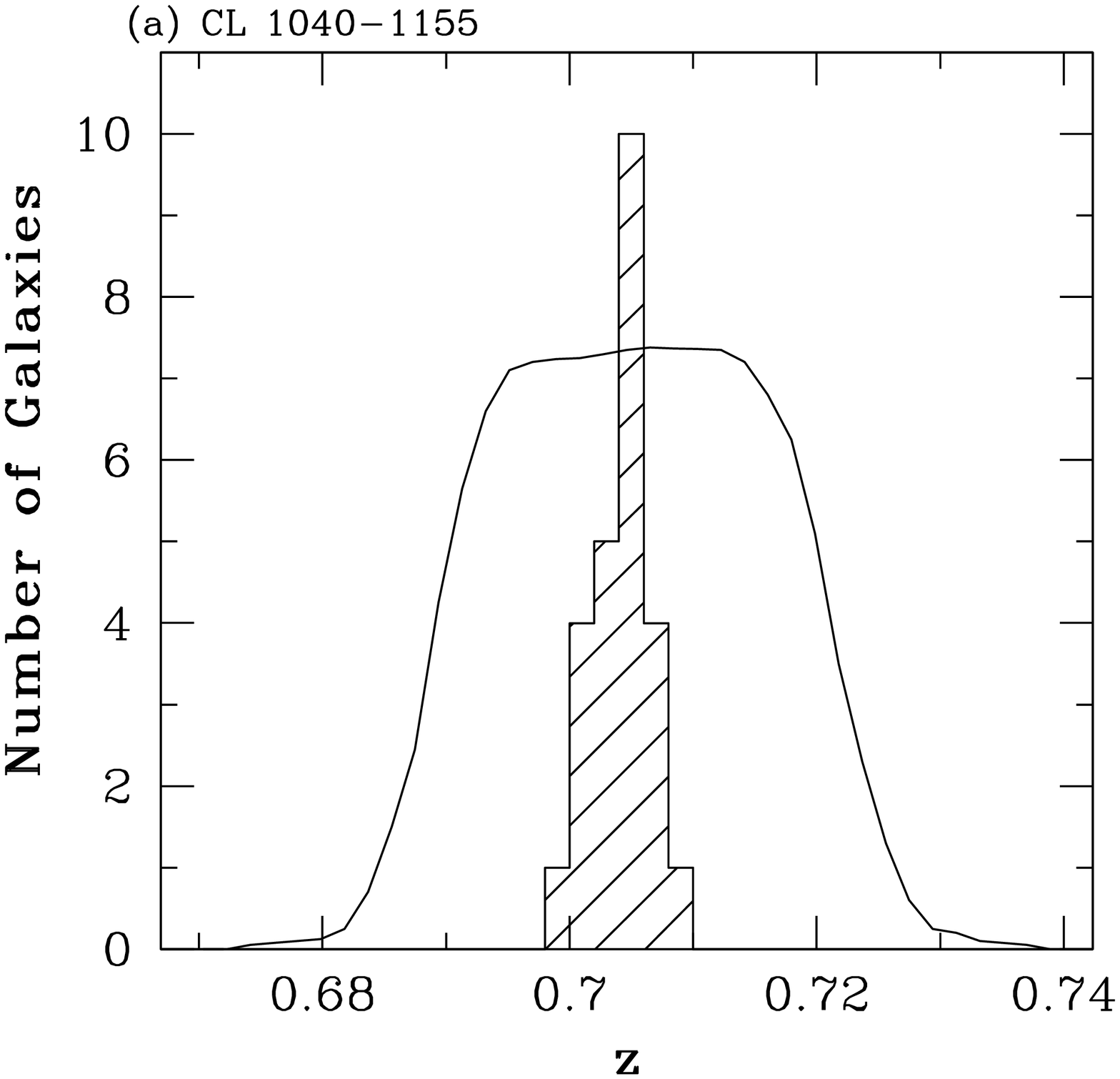}{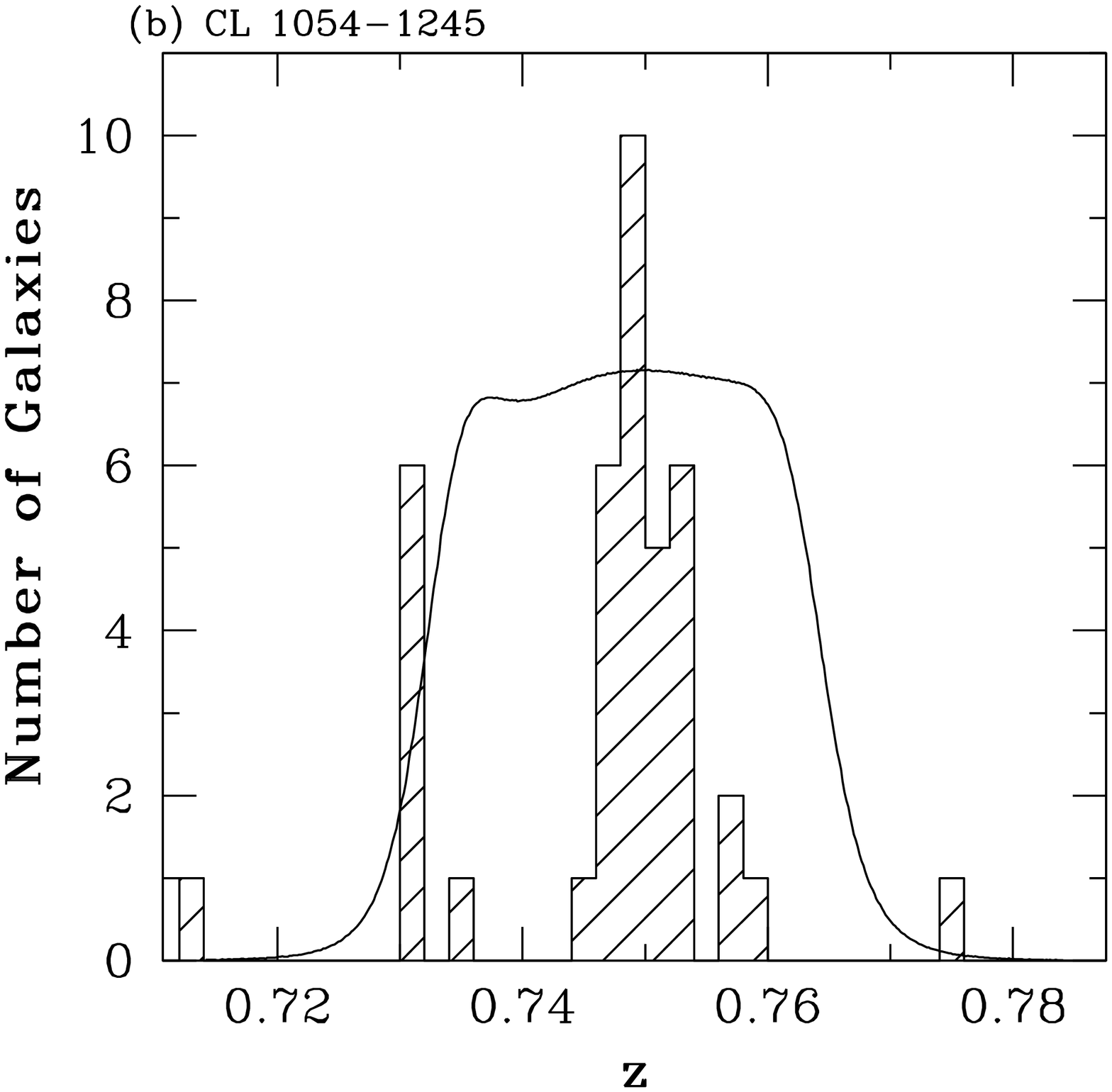}
\plottwo{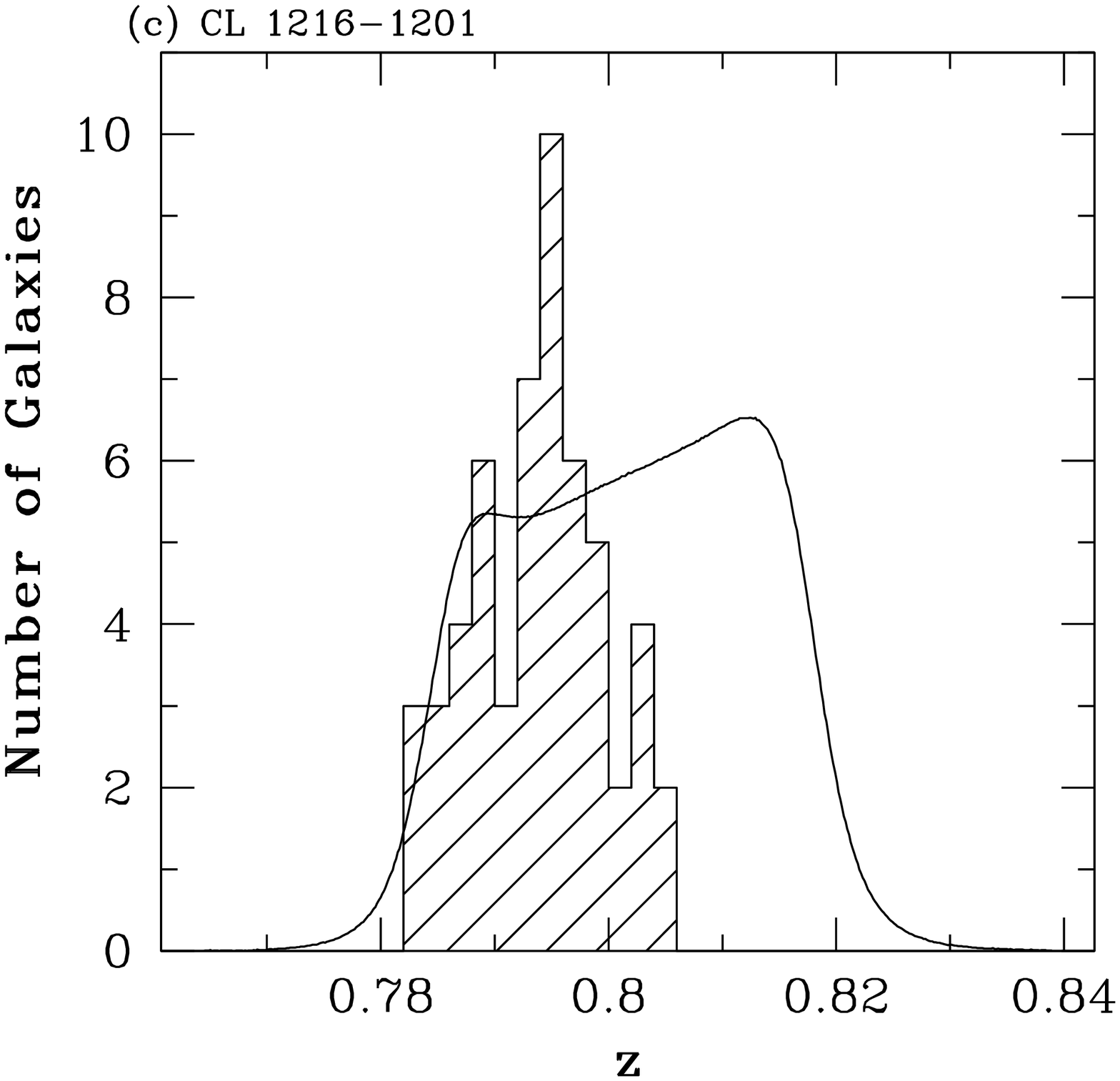}{f1d.ps}
\caption{Narrowband filter transmissions with histogram of spectroscopic
redshifts from \edi \ VLT survey for (a) \ca, (b) \cb, and (c) \cc.  Filter
transmissions are shown in units of \%/10 to match the vertical scale.
}
\label{comp_traces}
\end{figure}

The observations of the \ha\ narrowband flux are made with
custom narrowband (2\%) filters constructed by Barr Associates 
so that the central wavelength coincides with the
observed wavelength of \ha \ for each cluster's redshift.  The 2\% width
is well matched to the velocity dispersions of clusters. The 
filter transmission curves are plotted with results from VLT spectroscopy
({Halliday} {et~al.} 2004) in Figure \ref{comp_traces}.  
The mismatch in the central location of the 
filter for \cc\ is  due to a misleading preliminary redshift measurement of 
the cluster based on spectroscopy of a small number of 
members ({Nelson} {et~al.} 2001).
The integrated SFR for this cluster should be considered a lower-limit, 
although
judging from the Figure, the correction for missing galaxies is unlikely to 
exceed 10\%. Additionally, we note that the width of the filter used for \cj\
in Paper I is only 1\% (that filter was a stock narrowband filter).
Note that the 1\% filter is well-suited to the low velocity dispersion
of \cj \
($\sigma = $415~km/s;  {Postman} {et~al.} 1998).

We observe with the 
PISCES near-infrared camera ({McCarthy} {et~al.} 2001) and the 6.5m MMT.
PISCES has an inscribed, circular field-of-view with 
a 3.1\arcmin \ diameter and 0.18\arcsec/pixel.
The observations are summarized in Table \ref{comp_observations}.
The individual $J$-band exposure times range from 60 to 120 seconds, 
depending on sky brightness, and the individual narrowband
exposure times are ten minutes for \ca \ and \cb \
and five minutes for \cc.  The telescope is dithered
between successive images in increments of 10 to 15\arcsec \ 
in a $3 \times 3$ grid and is moved slightly between
successive grids.  
The total integration times are listed in Table \ref{comp_observations}.

Our data reduction procedure is described in detail in Paper I, 
and we only outline the procedure here.
Images are first corrected for cross-talk like contamination as described
in {McCarthy} {et~al.} (2001).  We then subtract a dark exposure.  
We flatten images 
using a sky flat created from the object frames.  
Using the IRAF contributed task DIMSUM, we make
a first-pass combined image 
and create an object mask from the combined image.
We then remake a sky flat with objects masked out.
We correct for geometric distortion using the IRAF task
GEOTRAN using a nearest interpolant and boundary extension. 
Images are then aligned, zeroed by the median sky level, 
and averaged with
pixels rejected according to the IRAF CCDCLIP algorithm.
The combined narrowband images show a residual peak-to-valley 
variation in flatness
at a level of $\le 2$\% across the entire image.  
We use SExtractor ({Bertin} \& {Arnouts} 1996) to create an image
of the background and then divide a normalized background image
into the narrowband frames.  The resulting images are flat
to within $\le 0.5$\% across the entire image.

\begin{figure}[h]
\plottwo{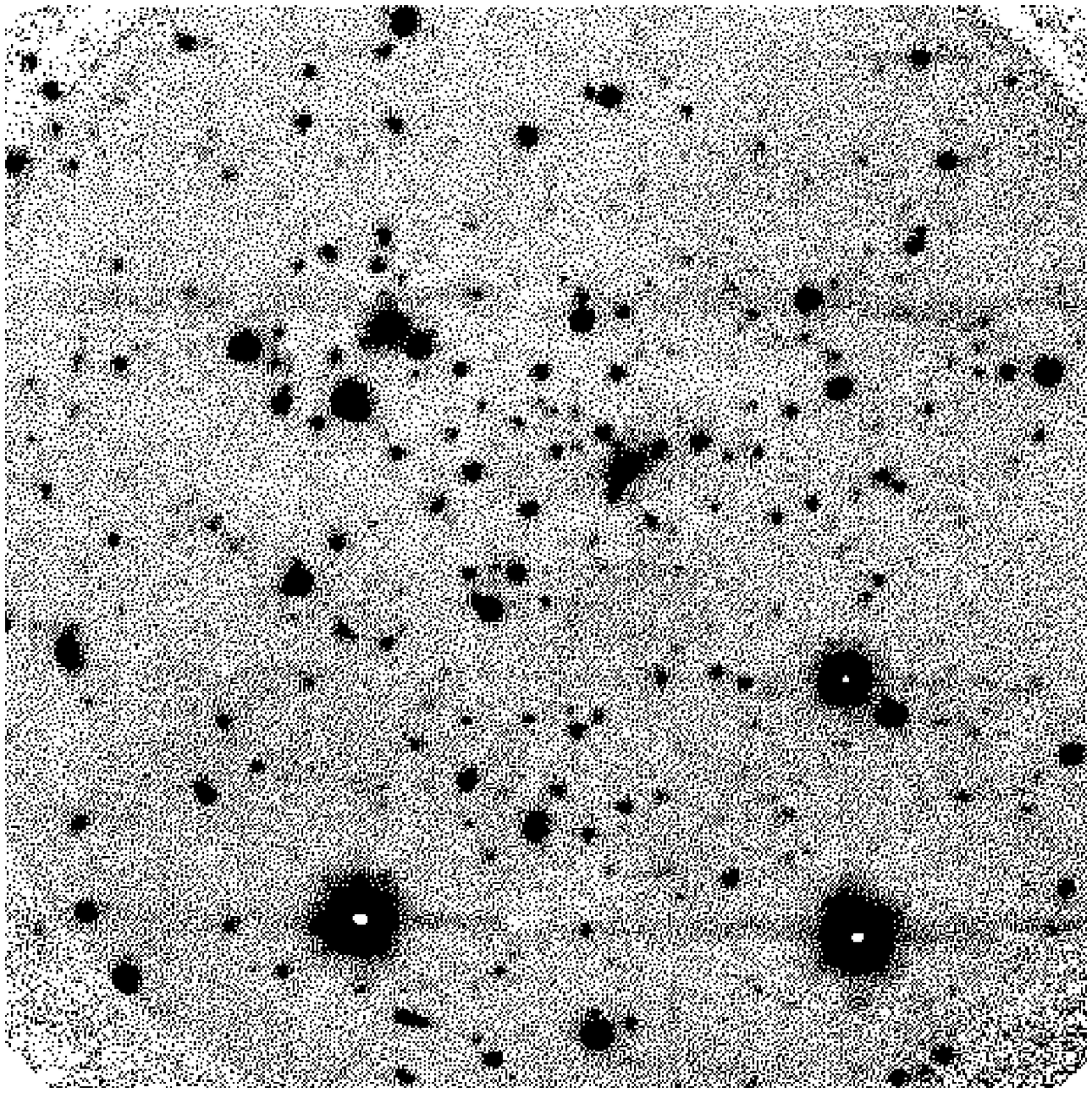}{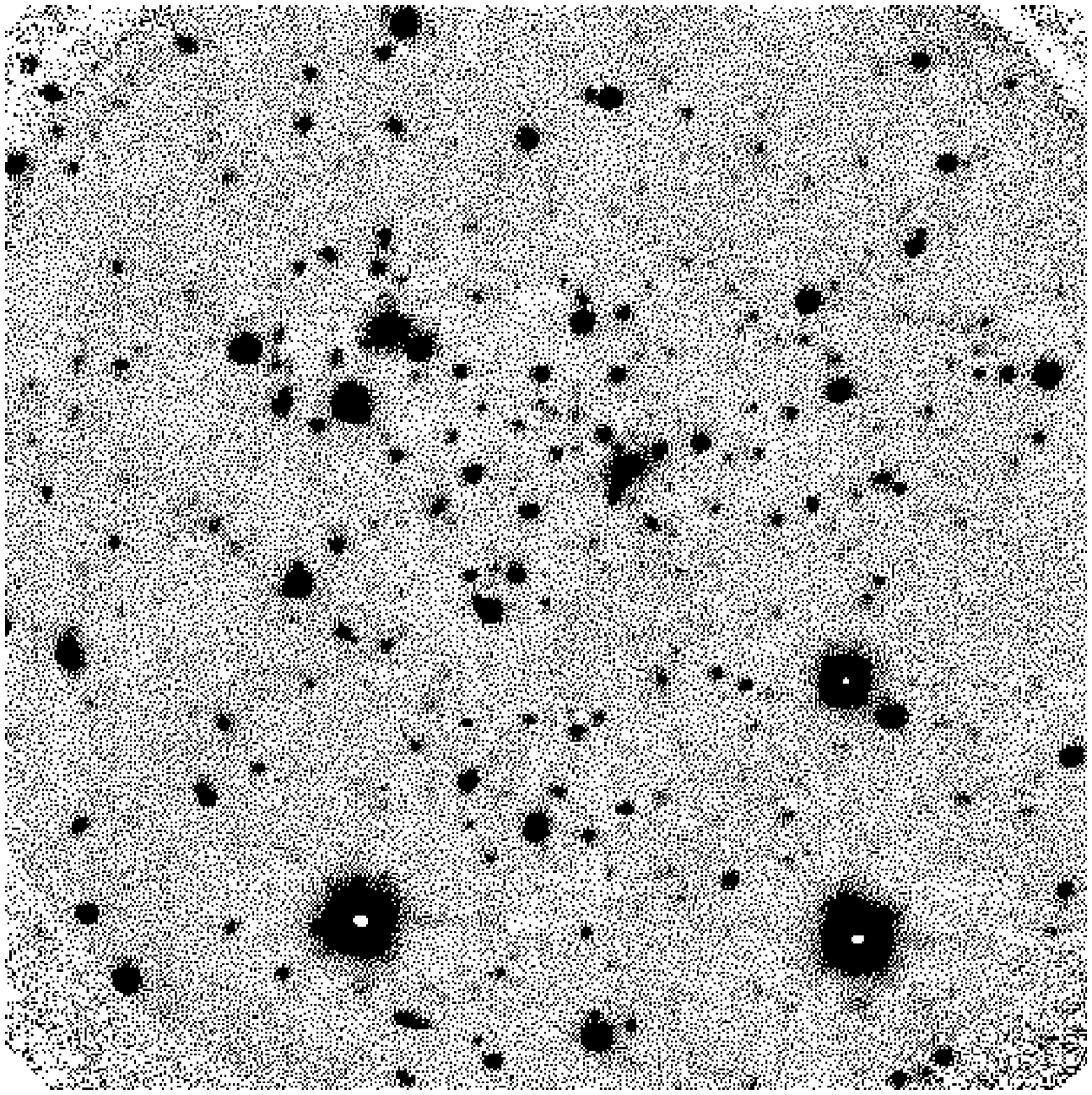}
\caption{(Left) Combined $J$-band image of \cb \ showing
horizontal streaking associated with bright stars.  (Right) Same
image after median filter is applied along rows and then subtracted from
original image.}
\label{streaking}
\end{figure}

The flattened images show horizontal streaking 
associated with bright stars that probably exists at some level
for all sources in the image.  The streaking, which severely
degrades the sky flatness, is not detectable in individual
images but is noticeable in the combined frame.  
We remove it by applying 
a median filter to the final
images, where the median filter has x-y dimensions of 150$\times$1 pixels.  
High and low thresholds are set to
reject any pixels slightly above or below the sky level
in the calculation of the median. This thresholding
rejects real objects while accepting the horizontal patterning.  
The filtered image is subtracted from the original,
and in Figure \ref{streaking} we
show the $J$-band image of \cb \ before (left) and after (right)
the median-filtered image is subtracted.
We use the unstreaked image to select objects and apertures, but measure
our photometry from the original images.

\subsection{Flux Calibration of $J$-band Images}
To flux calibrate the $J$-band data, we
 observe solar-type standard stars from  {Persson} {et~al.} (1998) 
in the $J$-band and narrowband filters.
We use the IRAF DAOPHOT package to solve the $J$-band 
photometric transformation, using a $J$ magnitude zeropoint for Vega of 
1600~Jy ({Campins} {et~al.} 1985) and solving only for the zeropoint 
and airmass terms.  
The solutions for 2002 December 19 and 2003 March 10 are listed in 
Table \ref{comp_phot}.

\begin{figure}
\plotone{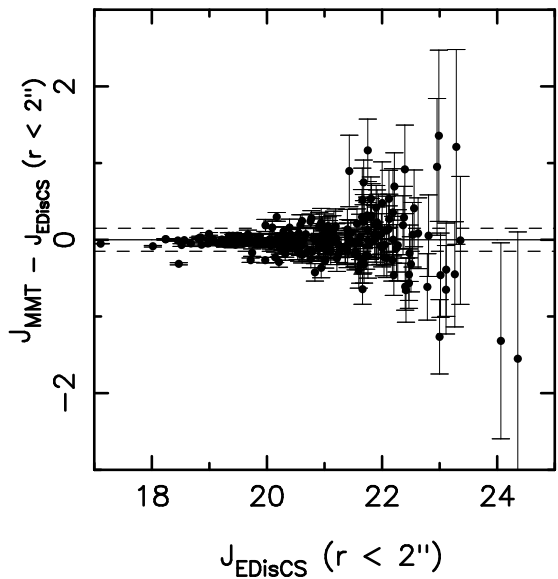}
\caption{Difference between MMT and \edi \ $J$ magnitudes measured within 
a 2\arcsec \ radius versus \edi \ $J$ magnitude for all likely cluster members
in the \ca, \cb,
and \cc \ fields.\label{compmmtedij}}
\end{figure}

The \edi \ collaboration has $J$ band data for the cluster
fields that are calibrated and analyzed independently from our data,
so we check to make sure there are no systematic zeropoint offsets
between the \edi \ and our $J$-band magnitudes.  
We limit our comparison to galaxies that are likely cluster members
based on photometric redshifts (\edi \ membership flag = 1; {Pell\'o} {et~al.} 2005),
and in Figure \ref{compmmtedij} we show the
difference in MMT and \edi \ magnitudes measured within 
a 2\arcsec \ radius versus \edi \ $J$ magnitude for all likely cluster members
that we detect in the \ca, \cb, and \cc \ fields.
We find an average difference in magnitudes for $J\rm_{EDisCS} < 23$ galaxies
of $0.01\pm 0.27$, $-0.02\pm 0.17$, and $0.07\pm 0.36$ for \ca, \cb, 
and \cc, respectively.  The average offset for \cc \ is
$0.02\pm 0.09$ if we restrict the comparison to $J < 20.5$ galaxies, so we 
do not correct for any zeropoint offset.

\subsection{Flux Calibration of Narrowband Images\label{nbfluxcal}}
\label{comp_fluxcal}
To flux calibrate the narrowband images we 
convert the $J$-band zeropoint to a narrowband (NB) zeropoint by 
correcting for differences in
(1) the bandwidth and
(2) atmospheric or filter/system transmission between the $J$ and NB filters.  
We refer to the product
of (1) and (2) as the filter throughput.

\begin{figure}
\plotone{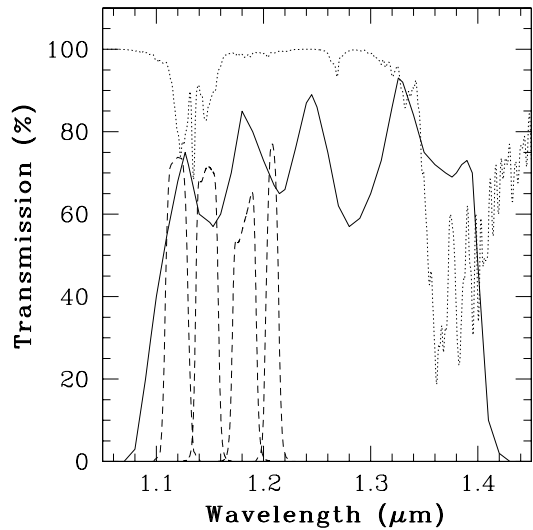}
\caption{Atmospheric transmission (dotted black line) plotted with $J$
(solid line) and narrowband (dashed line)
filter transmissions.  From left to right, the narrowband filters
correspond to \ca, \cb, \cc, and \cj.  
Atmospheric water feature at 1.10-1.15 \micron \
compromises flux calibration of \ca \ and \cb \
filters.}
\label{atrans}
\end{figure}

We calculate the relative throughputs of the $J$ and NB filters 
by multiplying the system and atmospheric transmission at each wavelength. 
The system transmission is the product of the filter transmission,
the mirror reflectivity, and the transmission of PISCES optics.
The atmospheric transmission varies during a night due to changing
levels of atmospheric water vapor.
Figure \ref{atrans} shows the atmospheric transmission (black
dotted line) as a function
of wavelength across the $J$-band window.  The filter transmission
for $J$ is shown with the solid line, and the NB filter 
transmissions are shown with dashed lines.
From left to right, the NB filters
correspond to \ca, \cb, \cc, and \cj \ from Paper I.
The time variability 
of the atmospheric water features at 1.10-1.15\micron \ and 1.35\micron, 
well documented by the 2MASS 
collaboration\footnote{http://www.ipac.caltech.edu/2mass/releases/allsky/doc/explsup.html}, 
complicates the flux calibration of the narrowband
images.  Because the water vapor can vary on short timescales (changing
the $J$ zeropoint by up to 0.1 through the course of 1 night according to 2MASS),
it is difficult to map its behavior, and this introduces uncertainty
into the calculated throughputs of the NB and $J$ windows.
We use two models of atmospheric transmission provided
by 2MASS for the Mt. Hopkins site, which correspond
to 0.5mm and 5.0mm of atmospheric water vapor.
We find that increasing the atmospheric water content
from  0.5mm and 5.0mm
translates into a 7.6$\pm$0.7\% variation in NB/J, where the standard
deviation reflects the differences among the three filters.
This systematic uncertainty is reflected in the narrowband zeropoints listed
in Table 1.

The goal of the narrowband imaging is to derive SFRs
for galaxies with excess NB flux, and to do so we must
convert the NB zeropoints to units of \smy. 
We first convert the NB zeropoint from Janskys to
$\rm erg~s^{-1}~cm^{-2}$ by multiply by the bandwidth of the NB
filter.
To obtain a luminosity from the measured flux 
we multiply
by $4\pi d_L^2$, where $d_L$ is the luminosity distance corresponding to
the cluster redshift.
We then use the Kennicutt star-formation relation ({Kennicutt} {et~al.} 1994) 
to convert \ha \ luminosity to SFR, where
\begin{equation}
\rm
1~ erg \ s^{-1} = 7.9 \times 10^{-42} \ M_\odot \ yr^{-1}.
\end{equation}
We correct for 1~magnitude of extinction at \ha \ ({Kennicutt} 1983), and we adopt
an [NII]/\ha \ ratio of 0.3 ({Tresse} {et~al.} 1999) 
to correct for [NII] contamination.
Using a large sample of local galaxies drawn from the SDSS, 
{Brinchmann} {et~al.} (2004) show that the Kennicutt SFR conversion is robust on 
average, although the conversion factor varies by $\sim$ 2.5 from the 
lowest to highest mass galaxies.  We do not have the additional spectral
information required to apply this first-order 
correction to the SFR conversion.
The average NB flux zeropoints for the 
\ca, \cb, and \cc \ filters are 
listed in Table \ref{comp_phot} in units of 
$\rm erg \ s^{-1} \ cm^{-2}$ and \smy.

\subsection{Source Detection and Photometry}
\label{comp_sex}
We use SExtractor for source detection and photometry.  We
detect all sources that are visually detected
using the following parameters:
a signal-to-noise threshold of 1.5$\sigma$ per pixel, a minimum object
area of 12 pixel$^2$, a 
tophat $5 \times 5$ convolution kernel, and a background mesh 
size of 48 pixels.
We run SExtractor in two-image mode, so that sources are selected 
from a combined $J$ and NB image, and the source positions
and apertures are then applied to the $J$ and NB images.
We make the NB+$J$ image by adding the sky-subtracted
NB image to the sky-subtracted, scaled $J$-band image,
where the $J$-band image is scaled by the ratio of NB-to-$J$ 
filter throughputs.  
We measure fluxes using isophotal apertures (FLUXISO) and do not apply
aperture corrections.

Sub-pixel dithering and geometric-distortion correction result
in correlated noise in neighboring pixels of our combined images, 
and thus Poisson noise models do not reflect the actual noise properties.
We empirically determine the noise properties of each final
$J$ and NB image following the method used by {Labb{\' e}} {et~al.} (2003).
First, we subtract a SExtractor BACKGROUND image to create
a sky-subtracted image.  We then select 500 random 
positions on each final image, avoiding detected objects and image edges,
and we measure the sky values in 15 circular apertures with radii 
ranging from 1 to 15 pixels.  We calculate the standard deviation in
sky values for each aperture size and then fit a two parameter noise model
where sky noise varies with the aperture linear size, $N = \sqrt{Area}$,
according to the following equation:
\begin{equation}
\sigma_{sky} = N  a (1 + b N).
\end{equation}
We let $a$ and $b$ vary between 0 and 1 in increments of 0.005 and 
select the $a$ and $b$ values that minimize the difference 
between predicted and measured sky noise.
We use these fits and an object's isophotal area to determine
the $J$ and NB isophotal flux errors.  
These errors are a factor of $\sim 1.5 - 2.5$ times
greater than the SExtractor isophotal flux errors.

The minimum object size and noise in the continuum-subtracted image
set our $1\sigma$ flux sensitivity.  
We calculate the 1 sigma noise associated with continuum-subtracted
flux in a 12 pixel$^2$ aperture using the noise models for
the $J$ and NB final images.  We list the $1\sigma$ noise  
in the last column on Table \ref{comp_observations} for the 
three \edi \ clusters.  
The detection thresholds correspond to $1 \sigma$ 
SFR limits of
0.18, 0.24, and 0.10 \smy \ for \ca, \cb, and \cc, respectively. 
The median sizes of galaxies in our final samples (see \S\ref{final}) are 
104, 92, and 92 pixel$^2$ for \ca, \cb, and \cc, 
which correspond to $1 \sigma$ SFR limits 
of 0.66, 0.83, and 0.32 \smy, respectively. 

\section{Continuum Subtraction }
\label{contsub}
\subsection{Estimating the Narrowband Continuum \label{nbcont}}
We estimate the continuum level in the NB filter from the
$J$-band flux levels, and this approach is complicated by two main issues.  
The first issue is the slope of an object's
spectral energy distribution (SED) through the $J$-band window.  
The second issue is the variability of atmospheric water vapor
and its effect on the transmission through the $J$ and 
NB filters.

\begin{figure}
\plotone{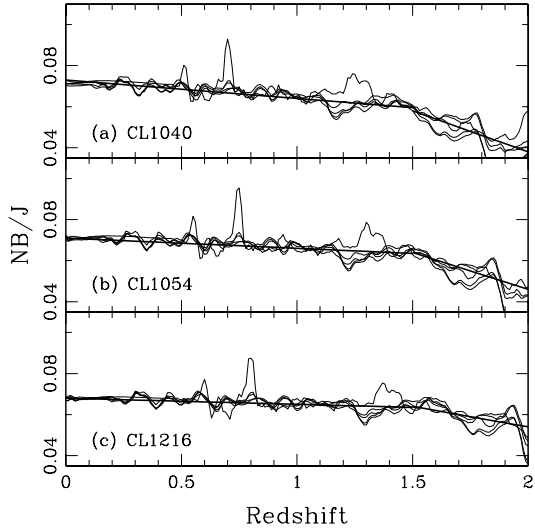}
\caption{Narrow-to-$J$ flux ratio for E, S0, Sa, Sb, and Sc galaxies (light lines)
as a function of redshift as observed through (a) \ca, (b) \cb, and 
(c) \cc \ filters.  
The bold black line shows the linear fit to NB/J versus
redshift, with a slope break at $z = 1.5$.
}
\label{ratioz}
\end{figure}

The scaled $J$-band flux gives a good estimate of the continuum in
the center of the $J$-band filter, but fails
at the blue or red end of the $J$ window because of 
the slope of a galaxy's SED.
As shown in Figure \ref{atrans}, the filters for the 
three \edi \ clusters lie in the blue end of the
$J$-band window.  
This results in a systematic variation in NB/J that is predominantly
a function of a galaxy's redshift.
To illustrate this, we show the narrow-to-J flux ratio (NB/J) 
for five galaxy types (light lines), E through Sc, as a function of redshift 
in Figure \ref{ratioz}.
The ratio as observed
through the \ca, \cb, and \cc \ NB filters are shown in panels 
(a), (b), and (c), respectively.
The galaxy SEDs are composite spectra from {Mannucci} {et~al.} (2001), and
the bold line shows a linear fit (by eye) to NB/J as a function
of redshift, with a break in slope at the rest-wavelength of 
4000\AA \ (approximately at $z = 1.5$).  
The redshift dependence of NB/J means that we can not 
estimate the continuum in the NB filter by simply scaling the
$J$-band flux by the ratio of filter throughputs, and 
a more sophisticated procedure for fitting the continuum is required.
An adjacent but non-overlapping NB filter
would provide a better estimate of the NB continuum, but this would
require a large increase in observing time.
The photometric redshifts from \edi \ collaboration ({Pell\'o} {et~al.} 2005) 
allow us to correct for the redshift
dependence of NB/J and therefore accurately estimate the NB continuum
from the $J$ band flux.
We use spectroscopic redshifts in \S\ref{spec}
to check the reliability of this technique.

\begin{figure}
\plotone{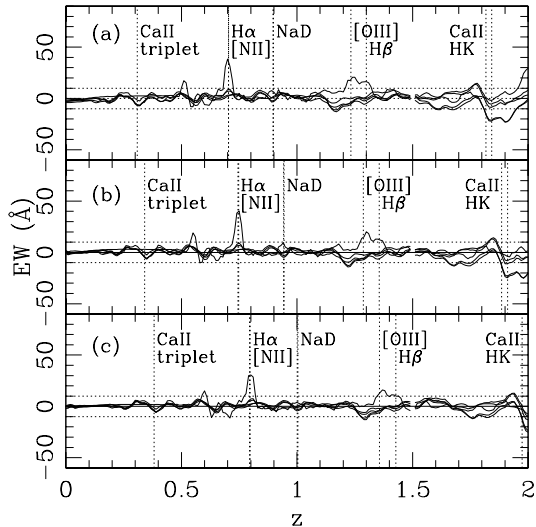}
\caption{EW for E, S0, Sa, Sb, and Sc SEDs
as a function of redshift, as observed through (a) \ca, (b) \cb, and 
(c) \cc \ filters after correcting for redshift dependence of 
NB/J using the bold lines in Figure \ref{ratioz}.
Dotted vertical lines show where NB filters can detect prominent 
spectral features 
other than \ha \ from non-cluster galaxies.  The unlabeled spectral features 
near $z \simeq 0.6$ are noise from the 
correction of telluric absorption at a rest-wavelength of $\sim$7400\AA.
Horizontal dotted lines show our minimum EW cut of 10\AA.
}
\label{galew}
\end{figure}

In Figure \ref{galew}, we show the observed equivalent width (EW) for the five galaxy
SEDs after we correct for the redshift dependence of NB/J using the bold lines
in Figure \ref{ratioz}.
Again, the results for the \ca, \cb, and \cc \ NB filters are shown
in panels (a), (b), and (c), respectively.
We define EW as
\begin{equation}
\label{eweqn}
EW = \frac{f_n - r f_J}{f_J} \Delta\lambda_J (\frac{1}{1+z_{cl}}),
\end{equation}
where $f_n$ is the NB flux in $\rm ADU~s^{-1}$,
$r$ is the calculated ratio of narrow-to-$J$ 
throughputs,
$f_J$ is the $J$-band flux in $\rm ADU~s^{-1}$, $\Delta\lambda_J$ is
the bandwidth of the $J$-band filter ($\equiv$ 0.25~\micron), 
and $z_{cl}$ is the cluster redshift.
With this definition, emission sources have positive EWs.
The residuals are largest for the \ca \ filter, where 
the average and standard deviation in EW for $z < 1.5$ galaxies 
are $-0.6 \pm 3.6$, $-0.08 \pm 3.3$, $0.2 \pm 3.6$, $1.6 \pm 3.1$, and
$4.2 \pm 6.9$\AA \ for the E, S0, Sa, Sb, and Sc spectra, respectively.  
The value for the Sc galaxy includes the \ha \ emission that
we are trying to detect, so
we overestimate the level of contamination. 
We conclude that given a rough measure of an object's redshift, 
we reliably measure EWs greater than 10\AA.
The unlabeled spectral features near $z \simeq 0.6$ are noise from the 
correction of telluric absorption at a rest-wavelength of $\sim$7400\AA.

Estimating the NB continuum from the $J$-band flux is also
complicated by the variability of atmospheric water vapor.
The calculated NB/J flux ratios depend on how much water vapor
is in the atmosphere because water directly affects the transmission
through the $J$ and NB filters.  
We can not precisely determine the water
content of the atmosphere because it varies over short timescales, 
which translates into a 7.6$\pm$0.7\% uncertainty in NB/J 
as discussed in \S\ref{nbfluxcal}.  To compensate,
we adjust the expected NB/J flux ratios within the range of predicted values
so that the peak of the observed
EW distribution is within $\pm5$\AA \ of zero for galaxies in our final sample.
For \ca, \cb, and \cc, we scale the J-band flux
by $0.0615\pm0.005$, $0.062\pm 0.005$, and $0.074\pm 0.005$ to estimate
the narrowband continuum,
where the errors reflect the uncertainty due to fluctuations in atmospheric
water vapor.
If all galaxies have some \ha \ emission then we will underestimate SFRs.
Our inability to definitively characterize the atmosphere introduces a 
systematic error of $\pm <$20\% in both EW and SFR.

\subsection{Quantifying Continuum-Subtracted Flux}
We discuss continuum-subtracted fluxes in terms of two quantities,
EW and SFR.  
When calculating EW, 
we assume all objects are at the cluster redshift, and
the uncertainty in EW is 
\begin{equation}
\sigma_{EW} = \frac{\Delta\lambda_J}{(1+z)} \sqrt{(\frac{1}{f_J})^2 \sigma_{f_{NB}}^2 + (\frac{f_{NB}}{f_J^2})^2 \sigma_{f_J}^2 + \sigma_{r}^2}.
\end{equation}
The NB and $J$-band flux errors, $\sigma_{f_{NB}}$ and $\sigma_{f_J}$, 
are the sum in quadrature of zeropoint
 and photometric errors.  The error in the narrow-to-$J$
ratio, $\sigma_r$, is the RMS of the ratio within the 68\% confidence intervals
of the photometric redshift.

The SFR is calculated by scaling the continuum-subtracted
flux by the conversion from ADU~s$^{-1}$ to \smy \ given in Table 
\ref{comp_phot}.  The continuum-subtracted flux, $f_{cs}$, is the scaled
$J$-band flux subtracted from the NB flux,
\begin{equation}
f_{cs} = f_{NB} - r f_J.
\end{equation}
The error in the continuum-subtracted flux is
\begin{equation}
\sigma_{fcs} = \sqrt{\sigma_{f_{NB}}^2 + r^2 \sigma_{f_J}^2},
\end{equation}
with errors in $\sigma_{f_{NB}}$, $\sigma_{f_J}$, and $\sigma_r$ defined as above.

\subsection{Comparison with Spectroscopy\label{spec}}
\begin{figure}
\plotone{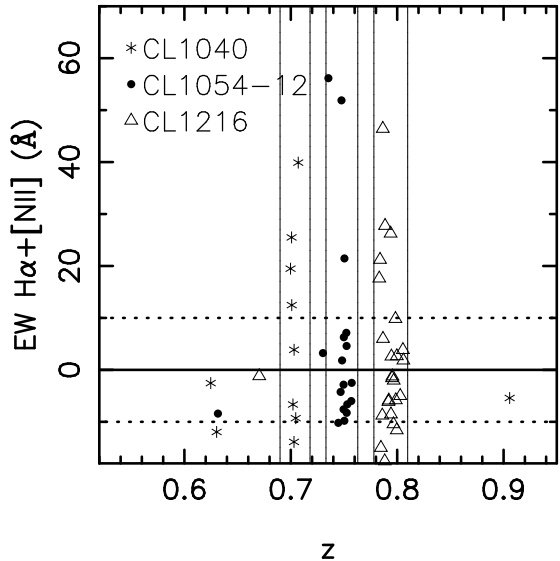}
\caption{\ewr \ versus spectroscopic redshift for all galaxies
that were targeted by \edi \ VLT spectroscopy.  
Horizontal dashed lines show our minimum \ewr \ cut of \ewmin.
Vertical solid lines bracketing cluster redshift show redshift range
where \ha \ falls in each NB filter.
All galaxies with significant emission are within $\Delta z = 0.02$ of the
cluster redshift.
\label{comp_haz}}
\end{figure}

The \edi \  collaboration has measured spectroscopic redshifts
for between 30 and 66 galaxies in each of these three clusters
({Halliday} {et~al.} 2004). We use the redshifts for all spectroscopic targets
that coincide with our \ha \ fields (we image only the central
2.5\arcmin$\times$2.5\arcmin)
to check that objects with significant NB emission
are actually at the cluster redshift.  In this section, we do {\bf not} use
photometric redshift or signal-to-noise of the continuum subtracted flux 
to restrict the galaxy sample.
In Figure \ref{comp_haz}, we plot \ha \ \ewr \ versus
spectroscopic redshift.  
The redshift range for which \ha \ is detectable through
each filter is shown with solid vertical lines that bracket 
each cluster redshift.  
The horizontal dotted lines show our minimum \ewr \ cut of \ewmin.
Figure \ref{comp_haz} illustrates the strength of \ha \ imaging
as an efficient method with low contamination 
for studying SFRs of high-redshift galaxies.

\begin{figure}
\plotone{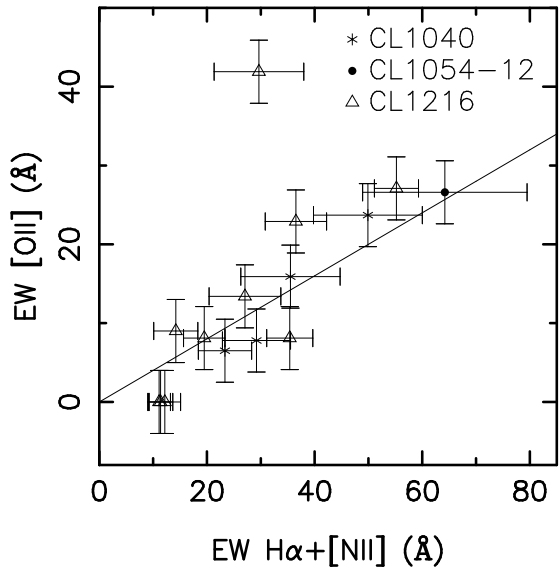}
\caption{Spectroscopically determined [OII] EW versus \ha \ \ewr \ 
from narrowband imaging.  Solid line shows empirical relation for local
galaxies (Kennicutt 1992a,b).
\label{ewhao2}}
\end{figure}

We compare spectroscopically determined [OII] EWs with our \ha \ EWs
as another check on our methods.  At low redshift, the EWs of [OII] and
\ha \ are well correlated, with $\rm EW([OII])~=~0.4 \ EW(H\alpha)$ 
(Kennicutt 1992a,b).
In Figure \ref{ewhao2} we show the correlation for the \edi \ galaxies
in our star-forming sample that have spectroscopically determined [OII].
The EWs are correlated with a hint of a steeper slope that needs to 
be confirmed with a larger sample.

\section{Results}
\label{results}
\subsection{Final Sample Selection \label{final}}
We use \edi \ photometric redshifts and spectroscopy to
select a final sample of cluster galaxies.
We first reject stars based on the \edi \ star flag,
which identifies stars using five-band photometry, object size,
and the SExtractor classifier index ({Pell\'o} {et~al.} 2005).  
We then include in our final galaxy sample all galaxies that are
likely cluster members based on photometric redshifts according to the
criteria of {Pell\'o} {et~al.} (2005).  
Specifically, the integrated probability 
that the photometric redshift is within 0.1 
of the cluster redshift
must be greater than 20\%.
The number of galaxies in each cluster that meet this selection criterion is:
\ca=38; \cb=63; \cc=134.
This is a purely photometric
selection and less stringent criteria were used to select spectroscopic
targets.  Comparison with spectroscopy indicates that 
less than 10\% of spectroscopically confirmed cluster members do not meet this 
photometric redshift criteria ({Pell\'o} {et~al.} 2005).
As a result, we also include spectroscopically confirmed members that did not
meet the photometric redshift cut (\ca=0; \cb=6; \cc=3; none of which 
has significant continuum-subtracted flux).
Finally, we include galaxies with significant continuum-subtracted
emission ($> 3 \sigma$ flux, \ewr\ $>10$ \AA) whose photometric redshift
68\% confidence interval includes the cluster redshift
(\ca=3; \cb=2; \cc=10). 
We omit any remaining galaxies that have significant narrowband emission
but photometric redshifts inconsistent with the cluster redshift
(\ca=7 ; \cb=5; \cc=3).  
A few galaxies (\ca=2; \cb=1; \cc=0) have no counterparts in the 
\edi \ catalogs, usually because they are near a bright star and thus confused
in the optical images from which the \edi \ catalogs are selected.
We do not include these objects in our samples.

The final samples include 41, 71, and 147 
galaxies for \ca, \cb, and \cc, respectively.
We compute the SFR and \ewr \ for each galaxy in the 
final sample assuming it is at the cluster redshift.
We present the data for our final samples in Tables 
\ref{cl1040}, \ref{cl1054}, and \ref{cl1216} for \ca,
\cb, and \cc, respectively.
The columns are described in the Table notes. 
We consider a galaxy to have significant \ha \ emission if it has
$>3\sigma$ continuum-subtracted flux and \ewr \ $>10$\AA.
We detect significant \ha \ emission for 10, 14, and 39 
galaxies in \ca, \cb, and \cc.  Hereafter, we refer to galaxies with
significant \ha \ emission as star-forming galaxies.

\begin{figure}
\epsscale{.65}
\plotone{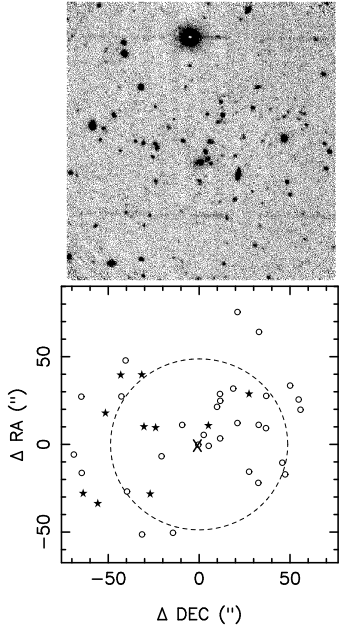}
\epsscale{0.605}
\hspace*{-.85cm}
\plotone{f9b.ps}
\caption{(Top) $J$-band image of \ca \ and (Bottom) 
schematic showing positions of galaxies in the final sample relative to cluster
center.
Galaxies with significant emission are marked with filled
stars.  Galaxies with no significant emission are shown
with open circles.
Image dimensions are 2.37\arcmin $\times$ 2.46\arcmin.
The dotted circle marks $0.5 \times R_{200}$, and the X
marks the position of the BCG.  We define \rtwo \ in \S\ref{intsfrs}.}
\label{cl1040jn}
\end{figure}

\begin{figure}
\epsscale{.65}
\plotone{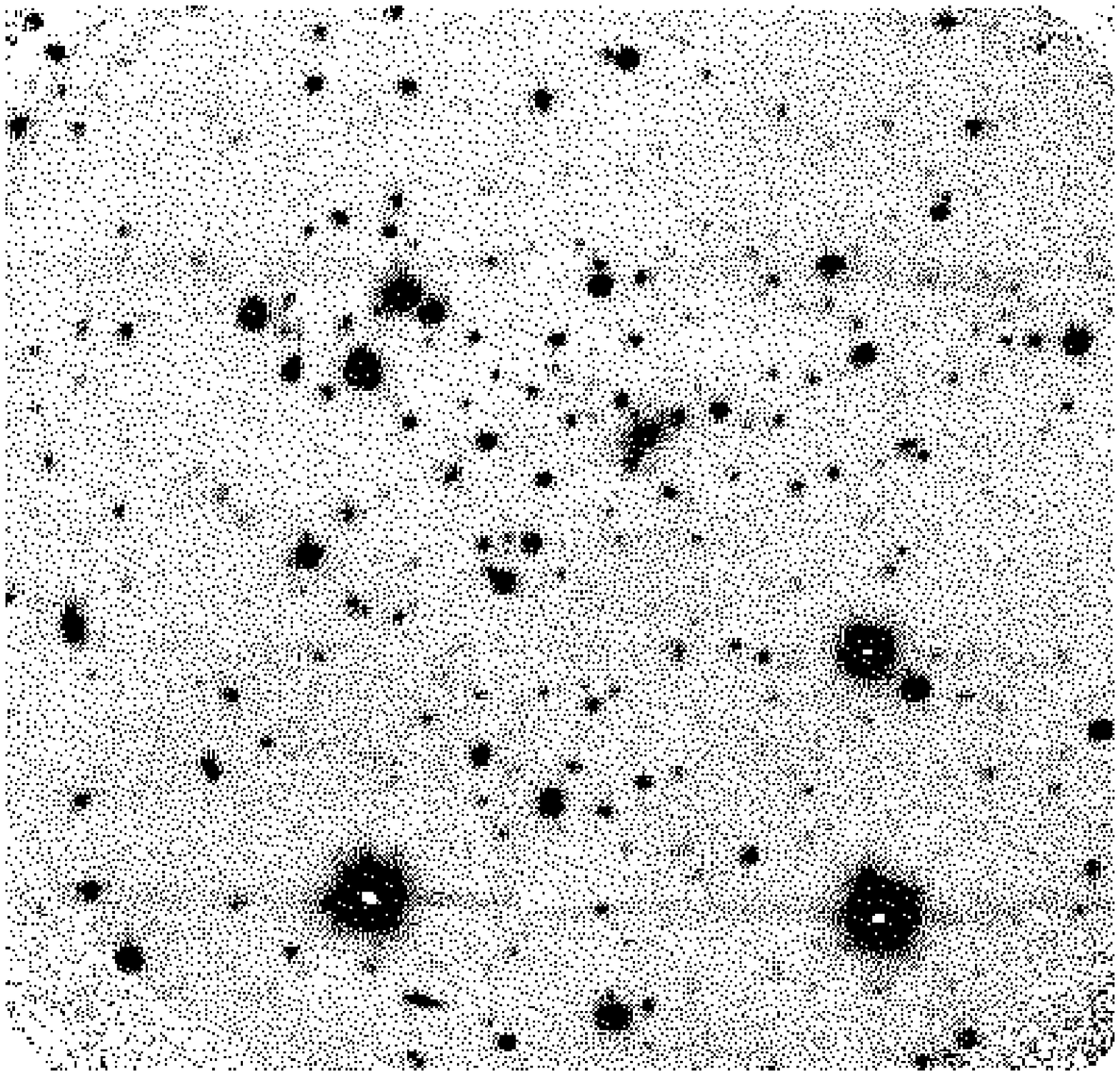}
\epsscale{0.605}
\hspace*{-.85cm}
\plotone{f10b.ps}
\caption{(Top) $J$-band image of \cb \ and (Bottom) 
schematic showing positions of galaxies in the final sample 
relative to cluster center.
Galaxies with significant emission are marked with filled
stars.  Galaxies with no significant emission are shown
with open circles.
Image dimensions are 2.73\arcmin $\times$ 2.62\arcmin.
The dotted circle marks $0.5 \times R_{200}$, and the X
marks the position of the BCG.}
\label{cl1054jn}
\end{figure}

\begin{figure}
\epsscale{.65}
\plotone{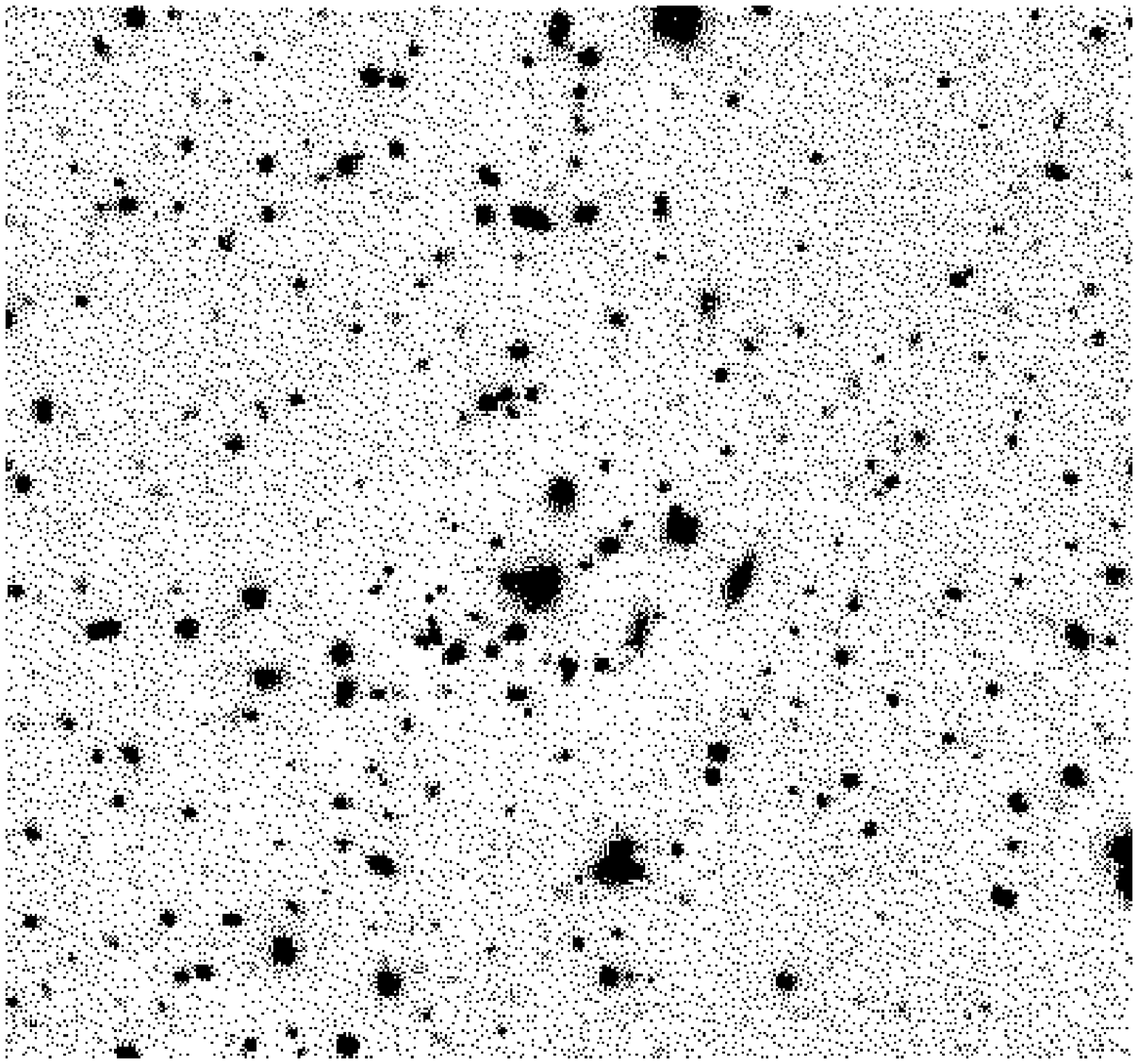}
\epsscale{0.605}
\hspace*{-.85cm}
\plotone{f11b.ps}
\caption{(Top) $J$-band image of \cc \ and (Bottom) 
schematic showing positions of galaxies in final sample relative to cluster
center.
Galaxies with significant emission are marked with filled
stars.  Galaxies with no significant emission are shown
with open circles.
The \cc \ image dimensions are 2.47\arcmin $\times$ 2.31\arcmin.
The dotted circle arc in the top right corner marks $0.5 \times R_{200}$, and the X
marks the position of the BCG.}
\label{cl1216jn}
\end{figure}

We show the positions of the final sample of 
galaxies with respect to the brightest cluster
galaxy (BCG) in the bottom panels of Figures \ref{cl1040jn}, \ref{cl1054jn}, 
and \ref{cl1216jn}, where
we represent star-forming galaxies with stars.
Galaxies with no significant NB emission are shown
with open circles.
The top panels show the final $J$-band images.
By inspection, the \cc \ field appears to be the richest cluster, 
and this is confirmed by velocity dispersion ({Halliday} {et~al.} 2004).  


\subsection{Properties of Star-Forming Galaxies \label{properties}}
\begin{figure}
\plotone{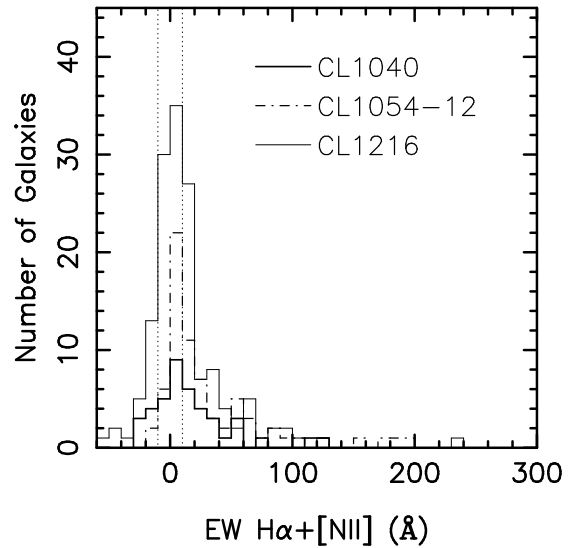}
\caption{Distribution of EWs for
all objects in final galaxy samples for \ca \ (dashed), \cb \ (dot-dashed) and \cc \ (solid).  
Vertical dotted lines show minimum reliable EW of $\pm$10~\AA.
}
\label{comp_histew}
\end{figure}

\begin{figure}
\plotone{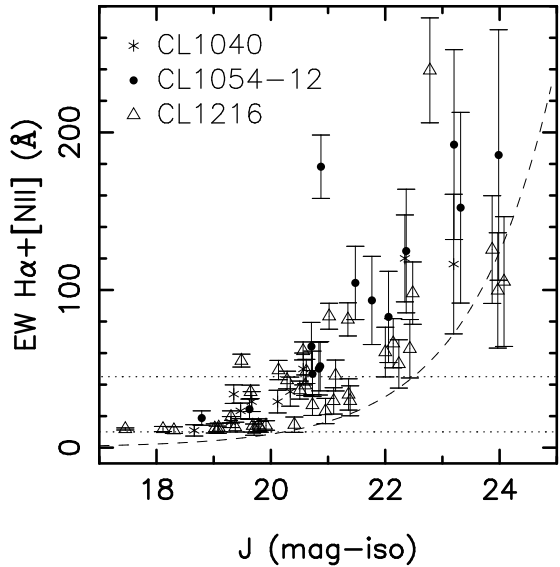}
\caption{$EW$ versus $J$ isophotal magnitude for all galaxies 
with $> 3\sigma$ 
continuum-subtracted flux and $EW > 10$\AA. 
The dashed curves show the approximate 
limits imposed by the $3 \sigma$ continuum-subtracted flux cut.\label{ewkall}
The horizontal lines show \ewr $= 10$ and 40\AA.
}
\end{figure}

We show the distribution of \ha \ \ewr s for the 
final samples of cluster galaxies
for \ca, \cb, and \cc \ in Figure \ref{comp_histew}.
None of the negative EWs is significant.
In Figure \ref{ewkall} we show \ewr \ versus $J$ isophotal magnitude 
for all star-forming galaxies.  The dashed curve shows the selection
imposed by our minimum continuum-subtracted flux cut.
The magnitude zeropoint, minimum SFR, and SFR conversion are different for
\ca, \cb, and \cc, so the curve is not the exact limit
for all three clusters but illustrates the
magnitude dependence of our selection criteria.
The horizontal dotted line at 10\AA \ shows our minimum \ewr \ cut,
and the line at 40\AA \ shows the \ewr \ cut used to define 
starburst galaxies in \S\ref{morphs}.
A population of faint galaxies with
low SFRs is beyond our detection limit, but nevertheless the range of EW increases at fainter
magnitudes.

\begin{figure}
\epsscale{1.}
\plotone{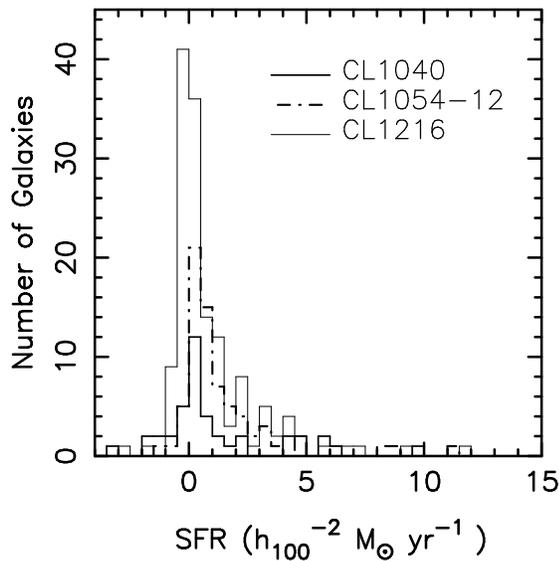}
\caption{Distribution of \ha-derived SFRs for
\ca \ (dot-dashed), \cb \ (dashed), and \cc \ (solid),
for all galaxies with EW $>$\ewmin.  Negative SFRs indicate objects with 
significant absorption.}
\label{complfi}
\end{figure}

\begin{figure}
\plotone{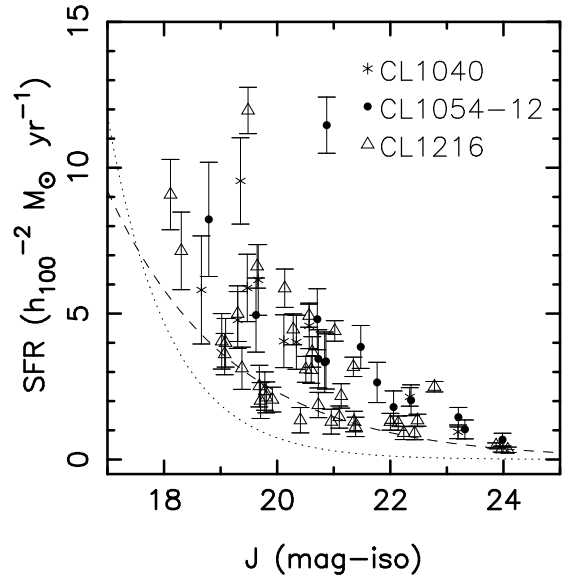}
\caption{SFR versus $J$ isophotal magnitude for all galaxies 
with $> 3\sigma$ 
continuum-subtracted flux and $EW > 10$\AA.   
The dashed curve shows the approximate limits imposed by the minimum flux cut,
and the dotted curve shows the effect of requiring $EW > 10$\AA.
\label{sfrkall}
}
\end{figure}

The distribution of \ha \ SFRs for all galaxies in the final sample 
is shown in Figure \ref{complfi}.
We show SFR versus $J$ isophotal magnitude for 
all star-forming galaxies in Figure \ref{sfrkall}.
The dashed curve shows the effect of our minimum flux limits, and the
dotted line shows the effect of our minimum \ewr \ cut.
We are more limited by our flux cut than by our minimum EW cut.
Again, the magnitude zeropoint, minimum SFR, and SFR conversion
are different for the three clusters, so the curves
do not show the exact selection used but illustrate the
magnitude dependence of our selection criteria.
The general trend is toward higher SFRs at brighter magnitudes. 
The bright galaxies are forming stars at a higher rate, 
but this is a smaller fraction relative to their overall stellar
population and hence their lower \ewr.  
The correlation between galaxy luminosity and SFR is 
also observed for field galaxies at similar redshifts (Tresse~et~al. 2002).

\begin{figure}
\plotone{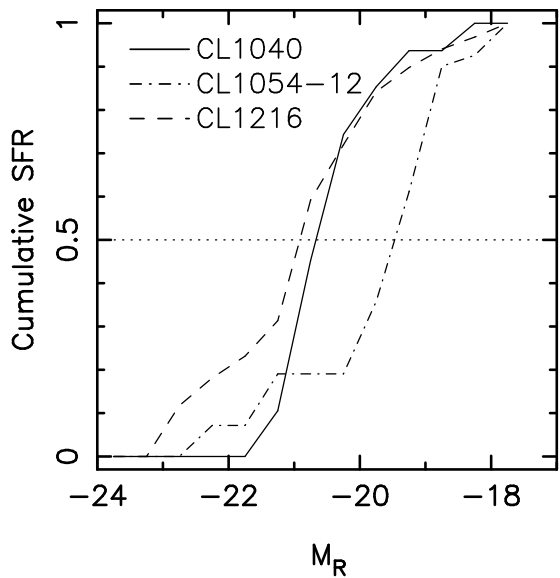}
\caption{Cumulative SFR versus $M_R$ magnitude for all 
star-forming galaxies. \label{cumsfrk}}
\end{figure}

In an effort to characterize the type of galaxies that
host the majority of star formation at $z \simeq 0.75$
and to provide a reference for lower redshift studies,
we calculate the cumulative SFR versus absolute $R$ magnitude ({Pell\'o} {et~al.} 2005).
We use 
deeper $J$-band imaging from the \edi \ collaboration 
to estimate completeness by comparing 
the number of galaxies per magnitude in 
our images with the number of galaxies
per magnitude in the \edi \ $J$-band images.  Magnitudes
are measured within 2\arcsec -radius apertures.
We multiply the total SFR in 
each magnitude bin by N(\edi)/N(MMT).  This likely overestimates our
incompleteness because we detect the most actively star-forming galaxies
at each magnitude and thus the missing galaxies should have lower SFRs. 
The resulting cumulative
distributions are shown in Figure \ref{cumsfrk}.
The 50\% mark occurs at $M_R \simeq -21$ for \cc \ and at 
$M_R \simeq -20.75$ for \ca.
The 50\% level occurs about one magnitude fainter in \cb, at $M_R \simeq -19.5$.   
The discontinuity in the cumulative distribution of \cb \ at $M_R \simeq -20$ 
is due to a galaxy that can be seen in 
Figures \ref{sfrkall} and \ref{ewkall} at $J = 20.93$, 
SFR~$= 11.5$~\smy, and \ewr $= 178$\AA.  
This galaxy accounts
for 24\% of the total SFR detected in \cb \ and may be an AGN.

\subsection{Environmental Variations in Star-Formation Properties}
\begin{figure}
\plotone{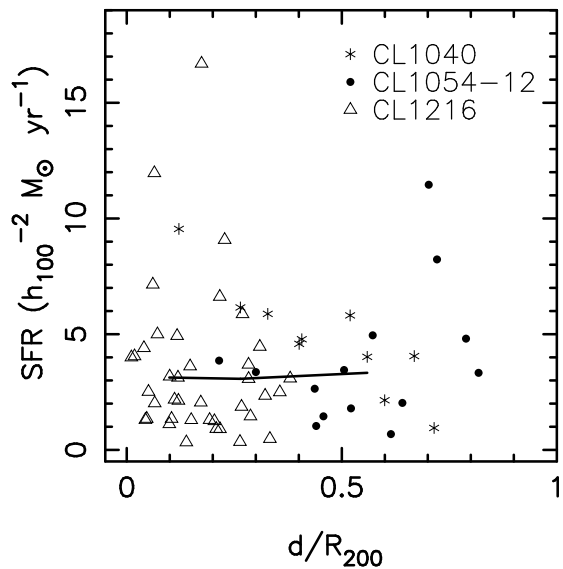}
\caption{SFR versus projected radial distance from the BCG
in terms of $R_{200}$ for all star-forming galaxies in final samples for \ca, \cb, and \cc.
The dashed vertical line shows where areal coverage
becomes incomplete.
The bold solid curve shows the median SFR 
versus projected radial distance.\label{sfrdall}}
\end{figure}

\begin{figure}
\plotone{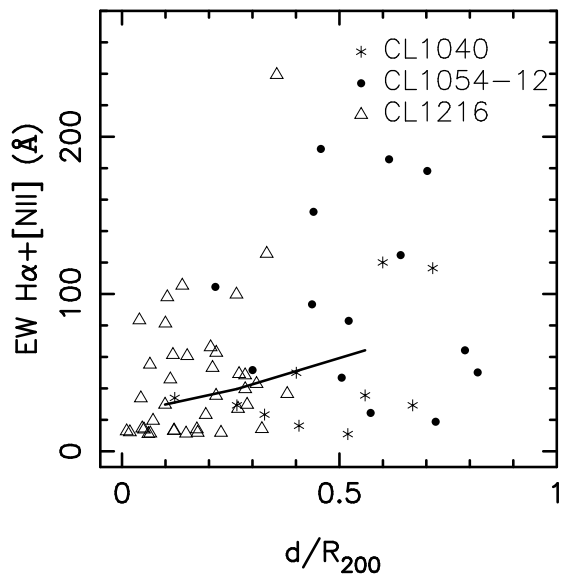}
\caption{\ewr \  versus projected radial distance from the BCG
in terms of $R_{200}$
for all star-forming galaxies in final samples for \ca, \cb, and \cc.
The dashed vertical line shows where areal coverage
becomes incomplete.  The bold solid curve shows the median \ewr \
versus projected radial distance.\label{ewdall}}
\end{figure}

For completeness we discuss the radial distribution of SFRs, but
our limited radial coverage of $r < 500$~\h~kpc precludes 
any definitive conclusions.  
In addition, any radial trends that exist will be 
weakened by projection effects.
To compare results for the three clusters, we express projected
radial distance in terms of $R_{200}$, which approximates the virial
radius and is described in detail in \S\ref{intsfrs}.
In Figure \ref{sfrdall} we plot the SFR versus projected radial distance 
from the cluster center for all star-forming galaxies in \ca, \cb, and \cc.
We use the brightest cluster galaxy to define the cluster 
center for the \edi \ clusters,
and we find no significant radial trend.  
Figure \ref{ewdall} shows  
\ewr \ versus projected radial distance for all star-forming galaxies.
The median \ewr \ of the star-forming galaxies, shown with the bold
line, increases with increasing projected radius.
A Spearman rank test indicates a 99.8\% probability that the two quantities
are correlated.  The interpretation of this trend is complicated by
the different detection limits and radial coverage 
for the three clusters.  The range of EWs
seems to increase with increasing radius, and this needs to be confirmed
with wider-field imaging.
We are in the processes of expanding 
our \ha \ imaging to larger radii in
order to track SFRs and \ewr\ from the cluster center to the field.  

\begin{figure}[h]
\plotone{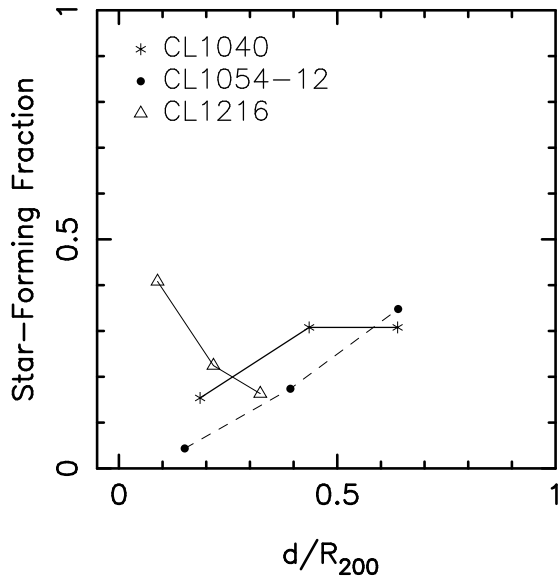}
\caption{Fraction of star-forming galaxies versus projected radial
distance from BCG in terms of $R_{200}$.
Points show star-forming fractions with galaxies 
in each cluster separated into 3 equally-populated bins.
\label{sffracd}}
\end{figure}

We show the fraction of star-forming galaxies versus projected distance
from the BCG in Figure \ref{sffracd}.  We divide the galaxies in each cluster
into three equally-populated bins and calculate the 
fraction of star-forming galaxies for each bin.  
\ca \ and \cb \ show
an increase in the star-forming fraction with increasing radius, but
\cc \ shows an anti-correlation.  Due to projection effects, 
we can not tell if these star-forming
galaxies at small projected radii are at the physical center of \cc.  

\begin{figure}[h]
\plotone{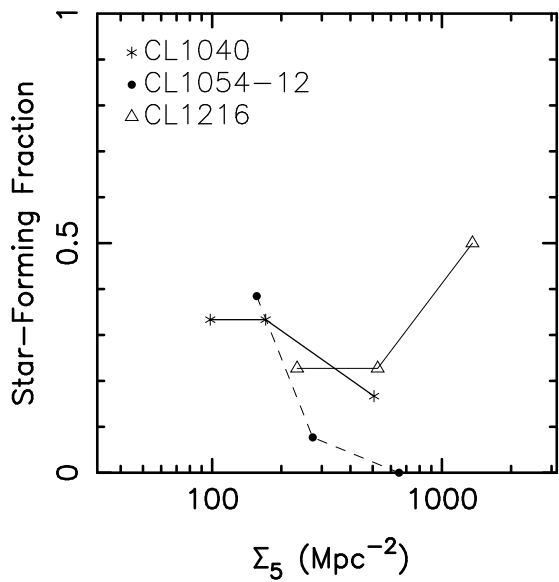}
\caption{Fraction of star-forming galaxies versus local galaxy surface
density as determined by 5th nearest neighbor.
Points show star-forming fractions for each cluster, with galaxies 
in each cluster separated into 3 equally-populated bins.
\label{sffracsigma10}}
\end{figure}

We examine SFR, EW and the fraction of star-forming galaxies 
as a function of local density.  
We characterize local density in terms of the surface density of galaxies, 
where
we use the distance to the 5th nearest neighbor to define the area in which
surface density is calculated.  We use only galaxies whose photometric 
redshifts are within 0.1 of the cluster redshift.  This corresponds to a 
larger redshift cut than the 1000~km/s cut used by Balogh~et~al. (2004).  As a
result, our local density estimates are not directly comparable to those
measured by Balogh et al.
We find no significant trend in either EW or SFR versus local galaxy 
surface density.  In Figure \ref{sffracsigma10} we show the fraction
of star-forming galaxies as a function of local galaxy density.  
The fraction  of star-forming galaxies in \ca \ and \cb \ decreases
with increasing surface density whereas \cc \ has the highest fraction
of star-forming galaxies at highest densities.  The results for \ca \ and
\cb \ are consistent with those of Balogh~et~al. (2004).  If we assume 
that star-forming galaxies are disk dominated, then the trends seen in 
\ca \ and \cb \ are also consistent with the results of 
Smith~et~al. (2004) and Postman~et~al. (2005), 
who demonstrate the existence of a morphology-density relation at $z \sim 1$.
Again, we are unable to tell whether the results for \cc \ reflect something 
physically different with this cluster or are due to projection effects.
 
\section{DISCUSSION}
\label{discussion}
\subsection{Comparison with Field Star-Formation Rates}
\label{comp_field}
The environmental dependence of star formation is 
now well-documented at low redshift.  The
2dF ({Lewis} {et~al.} 2002) and SDSS ({G{\' o}mez} {et~al.} 2003) studies 
use \ha \ emission to trace SFRs, and their results show
that the average SFR is lower in dense galaxy environments than in the field,
with the first signs of lower SFRs occurring in group environments.
{Balogh} {et~al.} (2004) show that this trend 
 is due to the
changing fraction of blue star-forming galaxies with environment,
where the lowest density environments contain 10-30\%
red galaxies and the cores of dense clusters contain $\sim$70\% red galaxies.
However, the origin of this trend remains debated, with the relevant
physical mechanisms that cause blue, star-forming galaxies to evolve into red, 
passive galaxies falling into three categories:
(1) starvation, where galaxies loose their extended gas halo
after entering a group or cluster environment and can no longer replenish their
star-forming fuel ({Larson} {et~al.} 1980; {Keres} {et~al.} 2004); 
(2) galaxy-galaxy interactions through which galaxies exhaust their fuel supply
in an interaction-induced burst of star formation while in the 
group environment, prior to merging with the cluster ({Zabludoff} \& {Mulchaey} 1998);
and (3) galaxy-inter galactic medium (ICM) interactions in which 
the cluster environment actively 
alters the star-forming properties of infalling galaxies through ram-pressure 
stripping ({Gunn} \& {Gott} 1972).
Several of these mechanisms are likely working together, with the dominant
physics changing with environment.  To date, astronomers
have not been able to quantify the relative importance of
starvation, galaxy-galaxy interactions, and galaxy-ICM interactions in
driving galaxies to evolve from blue to red.
Studying the environmental variations of SFRs at high-redshift provides
a necessary complement to low-redshift surveys
because galaxy-galaxy interaction rates and cluster
accretion rates 
were higher in the past ({van Dokkum} {et~al.} 1999; {Le F{\` e}vre} {et~al.} 2000), 
so their effects are more evident at high redshift than in the
local universe.

Current $z > 0.4$ cluster studies determine SFRs from 
spectroscopic measurements of the less robust
[OII]$\lambda 3727$ line.  
At intermediate redshift ($0.3 < z < 0.6$), the 
CNOC ({Balogh} {et~al.} 1997, 1998) and 
MORPHS ({Dressler} {et~al.} 1999; {Poggianti} {et~al.} 1999) surveys agree that
cluster galaxies of all Hubble types have lower SFRs than the
same type field galaxies. 
However, 
photometric modeling of the CNOC clusters favors a 
slow decline in star formation ({Ellingson} {et~al.} 2001; {Kodama} \& {Bower} 2001) as one approaches
the cluster, while
the MORPHS spectroscopy reveal a large population
of post-starburst cluster galaxies, which reflect sudden and
dramatic changes in SFRs.  
In a cluster at $z = 0.83$, {van Dokkum} {et~al.} (1999) 
find that although the observed merger rate
is significantly higher than the field, there is no sign of 
excess star-formation.  
{Postman} {et~al.} (1998, 2001) study four $z \sim 0.9$ clusters 
(including \cj \ from Paper~I)
and find that cluster galaxies have systematically lower
star-formation rates than field galaxies at similar redshifts.  
All of these studies agree
that cluster galaxies have lower SFRs than field 
galaxies, but they disagree about the likely cause.  
In addition, these studies are compromised by 
the unreliability of [OII]
emission as a SFR indicator.  

We take a first step in quantifying environmental variations in 
\ha-derived SFRs at $z \sim 0.8$ by comparing 
our cluster SFRs with field galaxy SFRs from the literature.
Two ground-based spectroscopic surveys provide \ha-derived 
SFRs for field galaxies in the same redshift range as our clusters.  
{Glazebrook} {et~al.} (1999) measure \ha \ fluxes
of thirteen galaxies drawn from the Canada-France 
Redshift Survey (CFRS).  
Galaxies are observed through a 1\arcsec \ slit, so they
apply an aperture correction of 1.7.  {Tresse} {et~al.} (2002) measure
the \ha \ flux for 33 CFRS field galaxies with redshifts between 0.5 and 1.1.  
They select galaxies with 
[OII] EW $>$10~\AA, which includes 78\% of $z > 0.5$ galaxies in the CFRS
sample.  They use a 2\arcsec \ slit width and 
conclude that no aperture correction
is required.  Both studies 
have high enough spectral resolution to resolve \ha \ and [NII], so 
their line fluxes are for \ha \ only.

To create a sample of field galaxies for comparison with our clusters,
we combine the two samples and limit the redshift range to 
$0.65 \le z \le 0.95$, which corresponds to  $\Delta t \sim \pm 1$~Gyr
relative to our cluster redshifts.  The combined sample includes
22 field galaxies between $0.7 \le z \le 0.93$, with
an average redshift of 0.816.
We calculate the SFR for each galaxy using the
published line fluxes and the SFR conversion described in 
\S\ref{comp_fluxcal}.
The minimum uncertainties associated with the {Tresse} {et~al.} (2002) and 
{Glazebrook} {et~al.} (1999) SFRs are $\sim$1~\smy.
Therefore, we apply a SFR threshold of $ 2$~\smy \ to the
field sample to minimize incompleteness.  This leaves 12 galaxies
in the field sample.
We apply the same selection to our cluster samples, and
we use bootstrap resampling to estimate the errors on
the median SFR for the clusters and field.  
The comparison between the cluster
and field galaxies is severely limited by the small size of the field sample.
We proceed with this caveat in mind.

\begin{figure}
\plotone{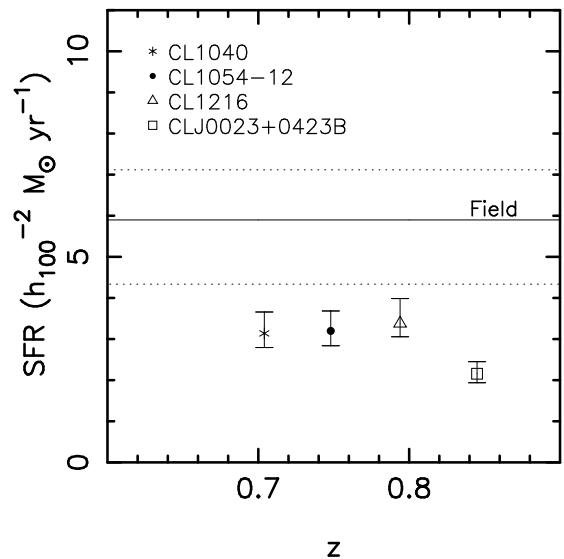}
\caption{Median cluster SFRs for all galaxies with SFRs $>2$~\smy \
versus cluster redshift.  Solid line shows median SFR of 12
$0.65 < z < 0.95$ field galaxies with same selection applied.  
Errorbars show $\pm 1 \sigma$
errors measured with bootstrap resampling.}
\label{comp_bootfield}
\end{figure}

We compare the median cluster and field SFRs 
for strongly star-forming galaxies in 
Figure \ref{comp_bootfield}.  The median field SFR
is shown with the solid line, and the $\pm 1\sigma$ 
bootstrap errors are shown with dotted lines.
The median SFRs for the clusters are shown as a function of cluster redshift, 
where again the $\pm 1\sigma$ errors are 
calculated with bootstrap resampling.
The uncertainty in the median field SFR is large given a 
sample of only 12 galaxies,
and we are not confident that the field surveys are complete 
above 2~\smy \ because the surveys do not quote completeness limits
in terms of SFR.  
If we apply a SFR cut at 3~\smy, the samples get 
uncomfortably small, but the median cluster SFRs remain systematically
below the field SFRs.  
The results suggest that the most actively star-forming field galaxies
are absent in the cluster environments.  This is different from
recent low-redshift study of {Balogh} {et~al.} (2004), where
they find that the distribution of \ha \ EWs for star-forming 
galaxies does not change with environment.  However, EWs and 
SFRs are not directly comparable, and a more direct comparison with
Balogh~et~al. results awaits a larger field sample.
We are in the process of using the same 
NB imaging technique to secure a 
sample of $>$500 $z \sim 0.8$ field galaxies whose SFRs and EWs will be 
directly comparable to our cluster sample.

Another issue complicating the field comparison is that we deliberately
biased the \edi \ cluster SFRs toward lower values by requiring the 
peak of the \ewr \ distribution to lie near \ewr $ = 0$\AA.  
This translates into
a $<20$\% systematic error in SFRs.  A systematic
increase of this magnitude for the cluster SFRs will 
bring the \edi \ SFRs within $1\sigma$ of the field value, 
but they still fall below the median field SFR.

\subsection{Evolution of Mass-Normalized SFRs\label{intsfrs}}
In this section, we characterize cluster SFR properties in terms of the
total SFR per cluster mass, \sfrm, and we compare
the \edi \ clusters and CLJ0023+0423B (Paper~I) with previous results
from the literature.
The advantages of this measure is that it is 
well-correlated with the fraction of emission
line galaxies ({Finn} {et~al.} 2005), yet
no correction for background/foreground galaxies is required
as is the case when calculating blue fraction or emission-line
fraction.

Defining uniform selection criteria is difficult when comparing 
spectroscopic and imaging studies, and comparing
results from [OII] and \ha \ is even more problematic.
Therefore, we limit our comparison to \ha \ studies of four 
clusters:  Abell 2390 at $z = 0.228$ ({Balogh} \& {Morris} 2000), 
AC~114 at $z = 0.32$ ({Couch} {et~al.} 2001), A~1689 at $z = 0.183$ 
({Balogh} {et~al.} 2002), and 
CL0024.0$+$1652 at $z = 0.4$ ({Kodama} {et~al.} 2004).
The {Balogh} \& {Morris} (2000) and {Kodama} {et~al.} (2004) studies employ
narrowband imaging to measure the \ha \ flux, 
whereas {Couch} {et~al.} (2001) and {Balogh} {et~al.} (2002) use spectroscopy.
We list properties of these clusters in Table \ref{cljlitsfrs}.

To calculate the total SFR for each cluster, we limit our analysis
to galaxies that lie within $0.5 \times R_{200}$ because 
the fraction of star-forming galaxies
is a strong function of the radial distance from the cluster center
for low-redshift clusters (e.g., {Balogh} {et~al.} 2004).
By definition, $R_{200}$, which approximates the 
virial radius, is the radius inside which the density is 
200 times the critical density:
\begin{equation}
\label{rtwoeqn}
200 \ \rho_c(z) = \frac{M_{cl}}{4/3 \pi R_{200}^3}.
\end{equation}
Using the redshift dependence of the critical density and the
virial mass to relate the line-of-sight velocity dispersion, $\sigma_x$,
to the cluster mass, we express $R_{200}$ as
\begin{equation}
R_{200} = 1.73 \ \frac{\sigma_x}{1000~{\rm km/s}} \ \frac{1}{
\sqrt{\Omega_\Lambda + \Omega_0 (1+z)^3}}\  h_{100}^{-1} \ {\rm Mpc}.
\end{equation} 
We choose a maximum radial extent of $0.5\times R_{200}$
to approximate the areal coverage of AC~114 
({Couch} {et~al.} 2001) and our higher redshift clusters.
The radial coverage of Abell 1689 does not extend to 
$0.5 \times R_{200}$.  We multiply the integrated SFR 
by 1.35 to correct for incomplete sampling within $0.5 \times R_{200}$, 
where we assume that
the galaxy distribution follows an isothermal sphere dark 
matter profile.  
This correction will still underestimate the integrated SFR
if there is a strong increasing radial gradient in either SFRs or the
fraction of star-forming galaxies.

When calculating the total SFR for the spectroscopic samples, 
we include all galaxies with velocities within $\pm 3 \sigma_x$.
The velocity sampling of the \cc \ NB filter
is comparable to  $\pm 3\sigma_x$ 
but is closer to $\pm 6 \sigma_x$ for the \ca \ and \cb \ filters.
Therefore, \ca \ and \cb \ may suffer from more contamination 
from nearby field galaxies.

Several other corrections are required when calculating the total
SFR for each cluster.
For the \edi \ clusters, we consider the SFRs of all galaxies in 
our final sample.  
We correct the spectroscopic surveys of
AC~114 and Abell~1689 for aperture bias and incomplete sampling 
by multiplying the integrated SFRs 
by 2.8, as suggested by {Kodama} {et~al.} (2004).
The imaging survey of {Balogh} \& {Morris} (2000) has
good areal coverage relative to $R_{200}$ but is sensitive to 
only the most actively star-forming galaxies with $\rm EW > 50$~\AA.  
If we apply the same \ewr \ cut to our \edi \ clusters, we would
detect 90, 24, and 63\% of the star-formation in 
\ca, \cb, and \cc.  We use the average completeness at 50\AA \ 
for the three \edi \ clusters (59$\pm$27\%) to estimate the completeness
of the Abell~2390 survey and therefore multiply the integrated
SFR of Abell~2390 by 1.7 to correct for star-formation missed 
from galaxies with $\rm EW < 50$\AA.

Some of the \ha \ emission we detect comes from active galactic
nuclei (AGN) and does not represent star formation.
Using spectra from the SDSS, 
{Kauffmann} {et~al.} (2003) find an AGN fraction of 10\% by number
including low and high luminosity AGNs, and the fraction of high
luminosity AGNs is 6\% in high density environments.
In a spectroscopic survey of a $z=0.83$ cluster, Homeier~et~al. (2005) 
find 2 AGN out of 102 confirmed cluster members.
We do not correct for AGN contamination, and we will be better able
to quantify contamination once the full EDisCS spectroscopic and X-ray 
results become available.

To calculate cluster mass, we use the virial mass to relate 
cluster mass, line-of-sight velocity dispersion, and $R_{200}$:
\begin{equation}
M_{cl} = \frac{3 \sigma_x^3 R_{200}}{G}.
\end{equation}
Combining with Equation \ref{rtwoeqn}, we express the cluster mass
solely in terms of velocity dispersion and cosmological parameters:
\begin{equation}
\label{mcl}
M_{cl} = 1.2 \times 10^{15} \ (\frac{\sigma_x}{1000 \ {\rm km/s}})^3\  \frac{1}{\sqrt{\Omega_\Lambda + \Omega_0 (1+z)^3}}
\ h_{100}^{-1} \ \rm M_\odot.
\end{equation}
Velocity dispersion is well correlated with 
cluster mass for relaxed clusters but will provide an overestimate
of cluster mass for clusters with substructure 
(C. Miller et al., private communication).
For this reason, we calculate the velocity dispersion for AC~114 and Abell~1689
from their X-ray luminosities using the best-fit $L_X - \sigma$ relation
of {Mahdavi} \& {Geller} (2001) because measured dispersions are inflated by
substructure.

\begin{figure}
\plotone{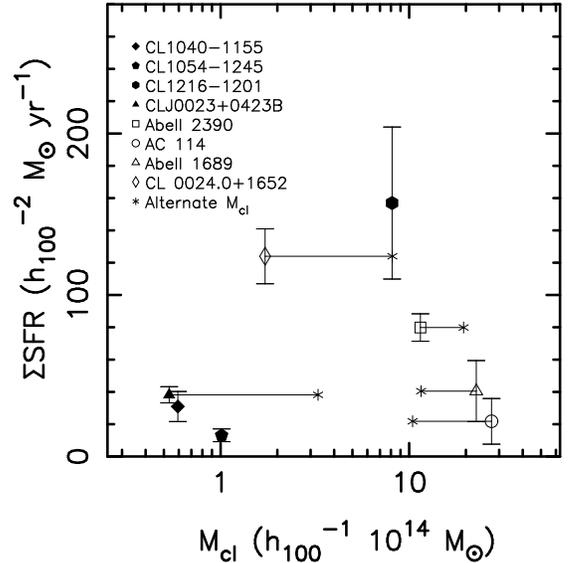}
\caption{Total SFR versus cluster mass for \edi \ clusters, \cj, and other 
\ha \ cluster surveys from the literature. Starred symbols show 
mass estimates derived from x-ray luminosities and/or lensing ({Kodama} {et~al.} 2004).
\label{sfrmcl}}
\end{figure}

In Figure \ref{sfrmcl} we 
first compare the total SFR to cluster mass, \mcl.
{Kodama} {et~al.} (2004) use other mass estimates derived from X-ray luminosities
and/or weak lensing for some of the clusters shown in Figure \ref{sfrmcl}, 
and we show these mass values with starred symbols.
Hereafter we use only the mass estimates
derived from Equation \ref{mcl} so that all clusters have masses determined 
from the same technique.  
We note that uncertainties in cluster
masses are a major source of error in this analysis because cluster
mass affects both the normalization and the estimate of $R_{200}$.

\begin{figure}
\plotone{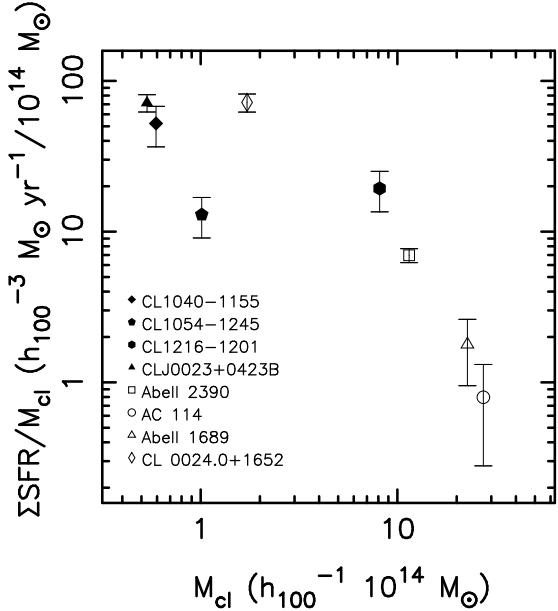}
\caption{Integrated cluster SFR per cluster mass versus cluster
mass for three the \edi \ clusters, \cj \ from Paper I, and lower redshift
clusters from the literature. 
\label{sfrmclmcl}}
\end{figure}

\begin{figure}
\plotone{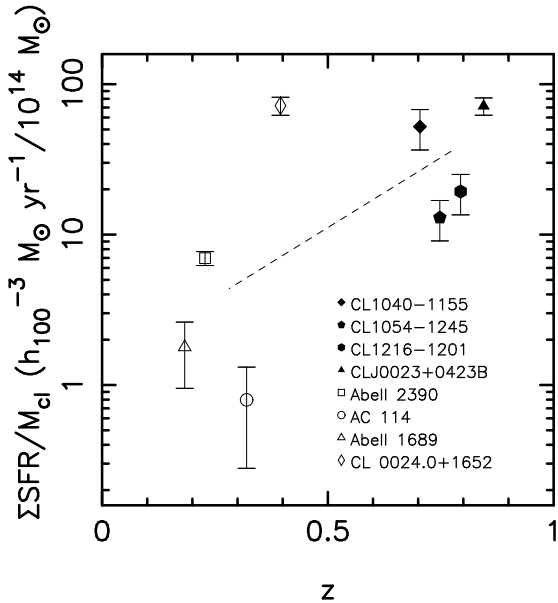}
\caption{Integrated cluster SFR per cluster mass versus cluster
redshift.  The dashed line shows the median increase in \sfrm \ 
between the $z \le 0.4$ and $z \simeq 0.75$ clusters. 
\label{sfrmclz}}
\end{figure}

We now normalize total SFR by cluster mass, and in Figure \ref{sfrmclmcl} we show
\sfrm \ versus
cluster mass. If galaxy SFRs and the fraction of star-forming galaxies were
independent of
cluster mass, then the data in this Figure would lie on a horizontal line.
Instead,
we find a strong relation between mass-normalized SFR and cluster mass that
is traced by 
both high and low redshift clusters.  This agrees with results from 
{Homeier} {et~al.} (2005), who find an anti-correlation  
between the mass-normalized SFR and cluster X-ray luminosity. 
This correlation complicates the classical
analysis of such data where one compares the SFRs across
redshift, as shown in Figure \ref{sfrmclz}. Interpreting
Figure \ref{sfrmclz} in light of Figure \ref{sfrmclmcl} suggests 
that much of what appears to be a strong
redshift dependence in mass-normalized SFR may be due to 
mass differences in the
clusters at the two redshifts. More directly, one would want to 
identify an offset in the relation
shown in Figure \ref{sfrmclmcl} for the two redshift 
epochs to identify an evolution in the SFRs.
However, these data are ill-suited for this comparison because the two
redshift samples
overlap minimally in cluster mass. Our full sample of 10 clusters, and a
complete analysis of SDSS data, will remedy this deficiency.

To demonstrate the potential for error, consider that the median increase
in \sfrm \ between the low and high redshift clusters is a factor of 8. It is
evident from Figure \ref{sfrmclmcl}, that over the mass range 
where the samples overlap, $\rm 10^{14} < M_{cl} < 10^{15}$, 
there is no evidence for any difference in the mass-normalized
SFR. This comparison is complicated by (1) SFRs determined
from NB imaging vs. spectroscopy, (2) different areal coverage of surveys,
(3) uncertainties in cluster mass determinations, and (4) inadequate 
overlap in cluster mass ranges.  To minimize the associated corrections, 
and ideal survey would select clusters that span the full range of 
redshift and cluster mass, measure SFRs within the same fraction of 
$R_{200}$ using the same technique, and have several
independent mass estimates for each cluster.

\subsection{Evolution of Starburst Galaxies\label{morphs}}
A goal of our survey is to help constrain the physical
mechanisms that cause blue, star-forming galaxies to evolve
into red, quiescent galaxies.  
Starburst galaxies are signposts of both 
galaxy-galaxy interactions (e.g., {Conselice} {et~al.} 2000; {Barton} {et~al.} 2000; {Homeier} \& {Gallagher} 2002) 
and ram-pressure induced bursts ({Bekki} \& {Couch} 2003) and so are
an important population for distinguishing among evolutionary scenarios.
We find a significant population of starburst (EW $> 40$\AA) galaxies in 
the $z \simeq 0.75$ clusters, and in this section we attempt
to track the evolution of these strongly star-forming galaxies
by combining our results with those from the MORPHS collaboration 
({Dressler} {et~al.} 1999; {Smail} {et~al.} 1997; {Poggianti} {et~al.} 1999).

The MORPHS survey consists of spectroscopy and ground and $HST$ WFPC2 
imaging of ten $0.35 < z < 0.5$ \ clusters.
Using the equivalent width of [O~II] and H$\delta$ 
to characterize star-formation properties,
{Poggianti} {et~al.} (1999) find a significant fraction of post-starburst
galaxies ($21 \pm 3$\%) in these clusters.
The redshift range of our sample is well suited to look
for the progenitors of this post-starburst population.  The look-back
time at the midpoint of our redshift range is about 2 Gyr larger than
at the midpoint of the MORPHS redshift range.
Spectral models of the post-starburst
galaxies show 
that star formation stopped between a few times $10^7$ and 1.5~Gyr
prior to the $z \simeq 0.45$ observations. If the star formation episode
lasts for at least a few hundred million years (the typical 
dynamical time of galaxy)
then the post-starburst galaxies at $z \simeq 0.45$ correspond to starburst
galaxies at $z \simeq 0.75$.

Although the relatively fortuitous timing suggests an interesting test, the
comparison is compromised somewhat by our lack of specific knowledge 
regarding the duration of both the starburst and post-starburst phases.
In addition, the MORPHS selections are 
based on [OII] emission and we use \ha \ emission.
Despite these complications, we proceed.
{Poggianti} {et~al.} (1999) find a post-starburst fraction of 21$\pm$3\% and a 
starburst fraction of 5$\pm$1\% for $M_V < -19 + 5\ {\rm log_{10}}\ h_{100}$. 
This magnitude cut brightens to $M_V < -19.52 + 5\ {\rm log_{10}}\ h_{100}$ when
we convert from the {Poggianti} {et~al.} cosmology ($q_0 = 0.5$) to the
one adopted in this paper ($\Omega_0 = 0.3, \ \Omega_\Lambda = 0.7$). 
In addition, we expect at least one magnitude of fading between 
the starburst and post-starburst phases ({Poggianti} {et~al.} 1999). 
Progenitors of the MORPHS post-starburst galaxies must 
therefore have $M_V < -20.52 + 5 \ {\rm log_{10}} \ h_{100}$
at $z \simeq 0.75$.

We calculate the starburst fraction for 
$M_V < -20.52 + 5 \ {\rm log_{10}}\ h_{100}$ \edi \ galaxies, where
the absolute V magnitudes for the \edi \ galaxies are
derived from the \edi \ photometry using the method of {Rudnick} {et~al.} (2003).
We define a starburst as a galaxy with 
EW(\ha)~$> 40$\AA \ ({Kennicutt} {et~al.} 1994; {Barbaro} \& {Poggianti} 1997) and limit the analysis
to galaxies within a projected radial
distance of $0.5 \times R_{200}$.  We calculate the starburst fraction
as the number of starburst galaxies divided by the total number of cluster
members within the same magnitude and radial cut.  We find a 
starburst fraction of 0$\pm$0\% (0/10) for \ca, 
0$\pm$0\% (0/13) for \cb, and 7$\pm$4\% (3/46) for \cc.
Combining statistics for the three clusters yields a starburst fraction of 
4$\pm$3\% (3/68).  We conclude that the observed
fraction of $M_V < -20.52 + 5 \ {\rm log_{10}}\ h_{100}$ starburst 
galaxies at $z \simeq 0.75$ can account
for the fraction of post-starbursts at $z \simeq 0.45$ 
if the post-starburst phase
lasts $\sim 5$ times longer than the starburst phase, consistent
with expected timescales.
We have neglected the effect of dust in our discussion,
and some starburst galaxies will have $\rm EW < 40$\AA \ due to 
selective dust extinction 
({Poggianti} {et~al.} 1999; {Poggianti} \& {Wu} 2000).
Even without taking dust into account,
we can explain the $z \simeq 0.45$ post-starburst
population with the $z \simeq 0.75$ starburst galaxies.

Some of the $EW > 40$\AA \ galaxies may be AGN.  Homeier~et~al. (2005)
find an AGN fraction of 2\% for a $z=0.83$ cluster.  If we assume
a similar fraction of AGN for the \edi \ clusters, then the starburst
fraction of $M_V < -20.52 + 5 \ {\rm log_{10}}\ h_{100}$ galaxies is 
reduced from 4\% to 2\%.  In this case, we can then account for the
$z \simeq 0.45$ post-starburst population if the post-starburst 
phase lasts 10 times longer than the starburst phase.

\section{Summary}
\label{summary}
We present \ha-derived star-formation rates for three $z \simeq 0.75$ 
galaxy clusters selected from the \edi \ survey. 
After combining the data from these three clusters
with that from Paper I, we conclude that:

\medskip
\noindent
1) The \ha\ NB imaging of clusters has low ($\sim 0\%$)
contamination and so is an efficient method with which to measure
ongoing star formation in high-redshift ($z \sim 0.8$) clusters.

\medskip
\noindent
2) We find no radial trend in SFRs among
the star-forming galaxies.  The median EW among star-forming galaxies 
increases slightly with radius, and the fraction of star-forming galaxies 
increases with radius in two out of three clusters.  
However, the radial coverage of our current imaging
is limited to $r < 500$~\h~kpc.

\medskip
\noindent
3) We find no trend in SFR or EW with local density, but we do find that
the fraction of star-forming galaxies decreases with increasing local density
in two out of three clusters..

\medskip
\noindent
4) Among star forming galaxies (SFR $> 2$~\smy), 
the median rate of star formation 
in cluster galaxies is less than that of field galaxies by $\sim$50\%.  
A larger sample of field galaxies is needed to confirm this result.

\medskip
\noindent
5) We characterize cluster evolution in terms of the mass-normalized 
integrated cluster SFR and find that the $z \simeq 0.75$ clusters 
have more SFR per cluster mass
than the $z \le 0.4$ clusters from the literature. 
The interpretation of this result is complicated
by the dependence of the mass-normalized SFR on cluster 
mass and the lack of sufficient
overlap in mass ranges covered by the low and high redshift samples.  

\medskip
\noindent
6) The fraction of starburst galaxies at $z \simeq 0.75$ is consistent 
with the fraction
of post-starburst galaxies at $z \simeq 0.45$ seen in the MORPHS clusters 
if the post-starburst phase lasts several ($\sim 5$) times
longer than the starburst phase.

\acknowledgements
RAF thanks John Moustakas, Marcia Rieke, and thesis committee 
members Rob Kennicutt and Chris Impey for useful
discussions regarding this work.
RAF acknowledges support from the NASA Graduate Student Researchers
Program through NASA Training Grant 
NGT5-50283
and from an NSF Astronomy and Astrophysics
Postdoctoral Fellowship under award AST-0301328.
DZ acknowledges support from the David and Lucile Packard Fellowship.
GR acknowledges the support of a
Goldberg fellowship at the National Optical Astronomy Observatory
(NOAO), which is operated by the Association of Universities for
Research in Astronomy (AURA), Inc., under a cooperative agreement with
the Natioanl Science Foundation.  GR also acknowledges the financial
support of the Sonderforschungsbereich 375 Astroteilchenphysik.
This research has made extensive use of the following: 
(1) NASA's Astrophysics Data System;  
(2) the 
NASA/IPAC Extragalactic
Database (NED) which is operated by the Jet Propulsion
Laboratory, California Institute of Technology, 
under contract with the National Aeronautics and Space Administration;
(3) online data catalogs provided by the Centre Donn\'ees 
astronomiques de Strasbourg; and (4) United States Naval Observatory star
catalog.



\begin{deluxetable}{lccccccc} 
\tablecaption{Summary of Observations \label{comp_observations} }
\tablehead{\colhead{Galaxy Cluster} &  \multicolumn{3}{c}{$J$-band} &\multicolumn{3}{c}{NB} & \\
\colhead{(1)} & \colhead{(2)} & \colhead{(3)} & \colhead{(4)} & \colhead{(5)} & \colhead{(6)} & \colhead{(7)} & \colhead{(8)} }
\startdata  
\ca	& 2002 Dec 19  	& 35	& 1.1 & 2002 Dec 19	& 150 & 1.13 & 0.18 \\
\cb	& 2003 Mar 10  	& 28	& 1.2 	& 2003 Mar 10	& 110 & 1.35 & 0.24\\
\cc	& 2003 Mar 11	& 36	& 1.0 	& 2003 Mar 12 	& 230 & 1.1\phn & 0.10\\
\enddata  
\tablecomments{Columns:  (1) Cluster name.  (2) Date of $J$-band observations.  (3) Exposure time of $J$-band image in minutes.  (4) FWHM of combined $J$-band image in units of arcsec.  (5) Date of NB image.  (6) Exposure time of NB in minutes.  (7)  FWHM of combined NB image in units of arcsec.  (8)  Continuum-subtracted flux $1\sigma$ detection limit in a 12 pixel$^2$ aperture in units of \smy.}
\end{deluxetable}

\begin{deluxetable}{llcccccc} 
\tablecaption{Summary of Photometric Calibrations \label{comp_phot}}
\tablehead{ & \multicolumn{3}{c}{$J$-band Solution} & \multicolumn{3}{c}{NB Flux ZP} \\
\colhead{Date} &\colhead{ZP} & \colhead{Airmass Coeff} & \colhead{R.M.S} &   
\colhead{Flux\tablenotemark{a}} & \colhead{SFR\tablenotemark{b}} & \colhead{Cluster} }  
\startdata  
{19 Dec 2002\tablenotemark{c}} 	&24.92$\pm$0.06 	& 0.06$\pm$0.05 & 0.0400 & 1.45$\pm$0.03 & 2.39$\pm$0.05 & \ca\\
\\
{10 Mar 2003\tablenotemark{d}} &24.93$\pm$0.07		& 0.08$\pm$0.05 & 0.0358 & 1.20$\pm$0.08 & 2.29$\pm$0.15 & \cb \\
	  &                       &               &                              & 0.94$\pm$0.06 & 2.10$\pm$0.14 & \cc\\
\enddata
\tablenotetext{a}{NB flux zeropoint in units of $\rm 10^{-16}~erg/ s/cm^{2}$.} 
\tablenotetext{b}{SFR in units of \smy \ corresponding to a $z = 0.704$, 0.748, and 0.794 
source for \ca, \cb, and \cc, respectively.} 
\tablenotetext{c}{Solution uses standard stars P525-E, P533-D, S840-F, and P545-C 
({Persson} {et~al.} 1998).}
\tablenotetext{d}{Solution uses standard stars P545-C, P550-E, P565-C, S791-C, S860-D, and S867-V ({Persson} {et~al.} 1998).} 
\end{deluxetable}


\clearpage
\begin{deluxetable}{rrrrrrrcrrc} 
\tablecaption{\ha \ Data for CL 1040-1155 Galaxies \label{cl1040}} 
\tablehead{\colhead{Name} & \colhead{$\delta$RA} & \colhead{$\delta$Dec} & \colhead{Flux$_n$} & \colhead{Flux$_J$} & \colhead{J} & \colhead{Cont. Sub} & \colhead{SNR} & \colhead{\ewr(\ha)} & \colhead{SFR} & \colhead{SF}\\  \colhead{(1)} &  \colhead{(2)} &  \colhead{(3)} &  \colhead{(4)} &  \colhead{(5)} &  \colhead{(6)} &  \colhead{(7)} &  \colhead{(8)} &  \colhead{(9)} & \colhead{(10)} & \colhead{(11)}} 
\startdata 
1040368-1156356 &    -51.4 &    -31.4 &      9.1 $\pm$      0.4 &    143.2 $\pm$      3.9 &    19.55 $\pm$     0.03 &     0.46 $\pm$     0.44 &      1.0 &   4.8 $\pm$   4.6 &   1.1 $\pm$   1.1 & 0 \\ 
1040368-1156185 &    -50.4 &    -14.4 &      0.6 $\pm$      0.1 &      9.2 $\pm$      1.3 &    22.53 $\pm$     0.14 &     0.08 $\pm$     0.14 &      0.6 &  13.2 $\pm$  23.5 &   0.2 $\pm$   0.3 & 0 \\ 
1040380-1157000 &    -33.8 &    -55.8 &      6.8 $\pm$      0.3 &     84.8 $\pm$      3.3 &    20.12 $\pm$     0.04 &     1.69 $\pm$     0.38 &      4.5 &  29.2 $\pm$   7.2 &   4.1 $\pm$   0.9 & 1 \\ 
1040383-1156311 &    -28.3 &    -26.9 &      5.3 $\pm$      0.3 &     56.1 $\pm$      2.9 &    20.57 $\pm$     0.05 &     1.91 $\pm$     0.32 &      5.9 &  49.9 $\pm$  10.1 &   4.6 $\pm$   0.8 & 1 \\ 
1040385-1156438 &    -26.9 &    -39.7 &      2.9 $\pm$      0.2 &     48.2 $\pm$      2.6 &    20.73 $\pm$     0.06 &     0.00 $\pm$     0.29 &      0.0 &   0.1 $\pm$   8.9 &   0.0 $\pm$   0.7 & 0 \\ 
1040384-1157079 &    -28.0 &    -63.8 &      0.7 $\pm$      0.1 &      5.0 $\pm$      0.9 &    23.20 $\pm$     0.18 &     0.40 $\pm$     0.10 &      4.1 & 116.5 $\pm$  44.3 &   0.9 $\pm$   0.2 & 1 \\ 
1040388-1155314 &    -21.9 &     32.5 &      0.9 $\pm$      0.1 &     14.1 $\pm$      1.5 &    22.06 $\pm$     0.11 &     0.09 $\pm$     0.16 &      0.6 &   9.3 $\pm$  17.1 &   0.2 $\pm$   0.4 & 0 \\ 
1040391-1155167 &    -17.1 &     47.3 &      2.6 $\pm$      0.2 &     56.0 $\pm$      2.6 &    20.57 $\pm$     0.05 &    -0.77 $\pm$     0.29 &      2.6 & -20.1 $\pm$   7.2 &  -1.8 $\pm$   0.7 & 0 \\
\enddata 
\tablecomments{Columns:  (1) Name is EDCSNJ followed by number listed in column.  (2) RA offset from BCG in arcseconds.  (3) DEC offset from BCG in arcseconds.  (4) Narrow-band flux in ADU/s. (5)  $J$-band flux in ADU/s. (6) J isophotal magnitude with SExtractor error.  (7) Continuum-subtracted flux in ADU/s.  (8) Signal-to-noise ratio of continuum-subtracted flux.  (9) Narrow-band \ewr \ in \AA.  (10) SFR in units of $h_{100}^{-2}\ \rm M_\odot \ yr^{-1}$.  (11) Star-forming galaxies that meet minimum continuum-subtract flux and \ewr \ cuts are denoted with 1. }
\end{deluxetable} 

\clearpage
\begin{deluxetable}{rrrrrrrcrrc} 
\tablecaption{\ha \ Data for CL 1054-1245 Galaxies \label{cl1054}} 
\tablehead{\colhead{Name} & \colhead{$\delta$RA} & \colhead{$\delta$Dec} & \colhead{Flux$_n$} & \colhead{Flux$_J$} & \colhead{J} & \colhead{Cont. Sub} & \colhead{SNR} & \colhead{\ewr(\ha)} & \colhead{SFR} & \colhead{SF}\\  \colhead{(1)} &  \colhead{(2)} &  \colhead{(3)} &  \colhead{(4)} &  \colhead{(5)} &  \colhead{(6)} &  \colhead{(7)} &  \colhead{(8)} &  \colhead{(9)} & \colhead{(10)} & \colhead{(11)}} 
\startdata 
1054378-1246245 &    -83.3 &    -33.0 &     10.1 $\pm$      0.8 &    165.6 $\pm$      4.3 &    19.33 $\pm$     0.03 &    -0.17 $\pm$     0.82 &      0.2 &  -1.4 $\pm$   7.1 &  -0.4 $\pm$   1.9 & 0 \\ 
1054378-1245513 &    -83.9 &      0.1 &      3.6 $\pm$      0.4 &     44.3 $\pm$      2.1 &    20.76 $\pm$     0.05 &     0.89 $\pm$     0.42 &      2.1 &  28.8 $\pm$  14.1 &   2.1 $\pm$   1.0 & 0 \\ 
1054383-1246113 &    -76.1 &    -19.9 &      1.4 $\pm$      0.3 &     17.4 $\pm$      1.4 &    21.78 $\pm$     0.08 &     0.34 $\pm$     0.27 &      1.2 &  28.1 $\pm$  23.4 &   0.8 $\pm$   0.6 & 0 \\ 
1054387-1245581 &    -70.0 &     -6.6 &      2.6 $\pm$      0.4 &     38.7 $\pm$      2.0 &    20.91 $\pm$     0.06 &     0.22 $\pm$     0.40 &      0.5 &   8.0 $\pm$  15.0 &   0.5 $\pm$   0.9 & 0 \\ 
1054388-1246514 &    -68.9 &    -60.2 &      4.0 $\pm$      0.4 &     41.3 $\pm$      2.3 &    20.84 $\pm$     0.06 &     1.45 $\pm$     0.45 &      3.2 &  50.1 $\pm$  16.8 &   3.3 $\pm$   1.0 & 1 \\ 
1054398-1246055 &    -54.3 &    -14.1 &     36.6 $\pm$      1.1 &    530.0 $\pm$      6.2 &    18.07 $\pm$     0.01 &     3.70 $\pm$     1.18 &      3.1 &  10.0 $\pm$   3.2 &   8.5 $\pm$   2.7 & 0 \\ 
1054393-1245361 &    -62.1 &     15.3 &     10.0 $\pm$      0.5 &    126.3 $\pm$      2.8 &    19.63 $\pm$     0.02 &     2.15 $\pm$     0.55 &      3.9 &  24.4 $\pm$   6.4 &   5.0 $\pm$   1.3 & 1 \\ 
1054391-1245439 &    -64.6 &      7.7 &      0.7 $\pm$      0.2 &      5.7 $\pm$      0.8 &    22.99 $\pm$     0.14 &     0.30 $\pm$     0.16 &      1.9 &  75.4 $\pm$  44.6 &   0.7 $\pm$   0.4 & 0 \\
\enddata 
\tablecomments{Columns:  (1) Name is EDCSNJ followed by number listed in column.  (2) RA offset from BCG in arcseconds.  (3) DEC offset from BCG in arcseconds.  (4) Narrow-band flux in ADU/s. (5)  $J$-band flux in ADU/s. (6) J isophotal magnitude with SExtractor error.  (7) Continuum-subtracted flux in ADU/s.  (8) Signal-to-noise ratio of continuum-subtracted flux.  (9) Narrow-band \ewr \ in \AA.  (10) SFR in units of $h_{100}^{-2}\ \rm M_\odot \ yr^{-1}$.  (11) Star-forming galaxies that meet minimum continuum-subtract flux and \ewr \ cuts are denoted with 1. }
\end{deluxetable} 

\clearpage
\begin{deluxetable}{rrrrrrrcrrc} 
\tablecaption{\ha \ Data for CL1216-1201 Galaxies \label{cl1216}} 
\tablehead{\colhead{Name} & \colhead{$\delta$RA} & \colhead{$\delta$Dec} & \colhead{Flux$_n$} & \colhead{Flux$_J$} & \colhead{J} & \colhead{Cont. Sub} & \colhead{SNR} & \colhead{\ewr(\ha)} & \colhead{SFR} & \colhead{SF}\\  \colhead{(1)} &  \colhead{(2)} &  \colhead{(3)} &  \colhead{(4)} &  \colhead{(5)} &  \colhead{(6)} &  \colhead{(7)} &  \colhead{(8)} &  \colhead{(9)} & \colhead{(10)} & \colhead{(11)}} 
\startdata 
1216420-1201509 &    -47.5 &    -33.6 &     27.2 $\pm$      0.5 &    368.6 $\pm$      3.0 &    18.46 $\pm$     0.01 &    -0.06 $\pm$     0.56 &      0.1 &  -0.2 $\pm$   2.1 &  -0.1 $\pm$   1.2 & 0 \\ 
1216418-1201081 &    -50.7 &      9.4 &     13.7 $\pm$      0.4 &    197.6 $\pm$      2.6 &    19.14 $\pm$     0.01 &    -0.88 $\pm$     0.46 &      1.9 &  -6.2 $\pm$   3.2 &  -1.8 $\pm$   1.0 & 0 \\ 
1216418-1200449 &    -51.4 &     32.3 &      9.9 $\pm$      0.3 &    126.4 $\pm$      2.0 &    19.63 $\pm$     0.02 &     0.57 $\pm$     0.33 &      1.7 &   6.3 $\pm$   3.7 &   1.2 $\pm$   0.7 & 0 \\ 
1216412-1201296 &    -58.7 &    -13.0 &      0.7 $\pm$      0.1 &     10.3 $\pm$      0.8 &    22.35 $\pm$     0.08 &    -0.07 $\pm$     0.12 &      0.6 &  -9.2 $\pm$  16.0 &  -0.1 $\pm$   0.3 & 0 \\ 
1216411-1201016 &    -60.7 &     15.6 &      0.1 $\pm$      0.0 &      1.4 $\pm$      0.3 &    24.53 $\pm$     0.20 &    -0.01 $\pm$     0.04 &      0.2 &  -7.5 $\pm$  36.2 &  -0.0 $\pm$   0.1 & 0 \\ 
1216428-1201073 &    -36.4 &     10.0 &    133.7 $\pm$      1.1 &   1685.3 $\pm$      5.5 &    16.81 $\pm$     0.00 &     9.27 $\pm$     1.14 &      8.1 &   7.7 $\pm$   0.9 &  19.5 $\pm$   2.4 & 0 \\ 
1216419-1202015 &    -50.0 &    -44.1 &      5.2 $\pm$      0.2 &     73.3 $\pm$      1.5 &    20.22 $\pm$     0.02 &    -0.21 $\pm$     0.25 &      0.9 &  -4.1 $\pm$   4.6 &  -0.4 $\pm$   0.5 & 0 \\ 
1216418-1201026 &    -51.1 &     14.5 &      0.8 $\pm$      0.1 &      9.5 $\pm$      0.7 &    22.44 $\pm$     0.07 &     0.08 $\pm$     0.10 &      0.8 &  11.3 $\pm$  14.6 &   0.2 $\pm$   0.2 & 0 \\
\enddata
\end{deluxetable}

\setcounter{table}{4}
\begin{deluxetable}{rrrrrrrcrrc} 
\tablecaption{ (Cont'd)  \ha \ Data for CL1216-1201 Galaxies}
\tablehead{\colhead{Name} & \colhead{$\delta$RA} & \colhead{$\delta$Dec} & \colhead{Flux$_n$} & \colhead{Flux$_J$} & \colhead{J} & \colhead{Cont. Sub} & \colhead{SNR} & \colhead{\ewr(\ha)} & \colhead{SFR} & \colhead{SF}\\  \colhead{(1)} &  \colhead{(2)} &  \colhead{(3)} &  \colhead{(4)} &  \colhead{(5)} &  \colhead{(6)} &  \colhead{(7)} &  \colhead{(8)} &  \colhead{(9)} & \colhead{(10)} & \colhead{(11)}} 
\startdata 1216453-1202177 &      0.7 &    -60.3 &      0.7 $\pm$      0.1 &     11.5 $\pm$      0.8 &    22.23 $\pm$     0.07 &    -0.12 $\pm$     0.11 &      1.1 & -14.8 $\pm$  13.0 &  -0.3 $\pm$   0.2 & 0 \\ 
1216452-1202262 &     -0.1 &    -68.6 &      5.3 $\pm$      0.2 &     72.8 $\pm$      1.3 &    20.23 $\pm$     0.02 &    -0.09 $\pm$     0.21 &      0.4 &  -1.7 $\pm$   4.0 &  -0.2 $\pm$   0.4 & 0 \\ 
1216453-1201300 &      0.2 &    -12.8 &      0.8 $\pm$      0.1 &     14.4 $\pm$      0.7 &    21.99 $\pm$     0.05 &    -0.25 $\pm$     0.11 &      2.3 & -24.1 $\pm$   9.8 &  -0.5 $\pm$   0.2 & 0 \\ 
1216451-1200359 &     -1.7 &     41.2 &      1.8 $\pm$      0.1 &     26.2 $\pm$      1.1 &    21.33 $\pm$     0.04 &    -0.16 $\pm$     0.17 &      0.9 &  -8.5 $\pm$   8.7 &  -0.3 $\pm$   0.4 & 0 \\ 
1216452-1201317 &     -1.0 &    -14.5 &      1.0 $\pm$      0.1 &     15.2 $\pm$      0.8 &    21.93 $\pm$     0.06 &    -0.10 $\pm$     0.12 &      0.8 &  -8.8 $\pm$  10.9 &  -0.2 $\pm$   0.3 & 0 \\ 
1216452-1202231 &     -1.2 &    -65.7 &      0.3 $\pm$      0.0 &      2.4 $\pm$      0.3 &    23.95 $\pm$     0.14 &     0.11 $\pm$     0.04 &      2.5 &  65.8 $\pm$  32.3 &   0.2 $\pm$   0.1 & 0 \\ 
1216494-1200189 &     61.2 &     58.2 &      0.7 $\pm$      0.1 &     11.1 $\pm$      0.7 &    22.26 $\pm$     0.06 &    -0.08 $\pm$     0.10 &      0.8 &  -9.8 $\pm$  11.9 &  -0.2 $\pm$   0.2 & 0 \\ 
1216450-1201253 &     -3.4 &     -8.7 &      2.5 $\pm$      0.1 &     25.6 $\pm$      1.1 &    21.36 $\pm$     0.05 &     0.62 $\pm$     0.17 &      3.7 &  33.7 $\pm$   9.9 &   1.3 $\pm$   0.4 & 1 \\
\enddata 
\tablecomments{Columns:  (1) Name is EDCSNJ followed by number listed in column.  (2) RA offset from BCG in arcseconds.  (3) DEC offset from BCG in arcseconds.  (4) Narrow-band flux in ADU/s. (5)  $J$-band flux in ADU/s. (6) J isophotal magnitude with SExtractor error.  (7) Continuum-subtracted flux in ADU/s.  (8) Signal-to-noise ratio of continuum-subtracted flux.  (9) Narrow-band \ewr \ in \AA.  (10) SFR in units of $h_{100}^{-2}\ \rm M_\odot \ yr^{-1}$.  (11) Star-forming galaxies that meet minimum continuum-subtract flux and \ewr \ cuts are denoted with 1. }
\end{deluxetable}

\clearpage
\begin{deluxetable}{lccccccccccc} 
\tablecaption{Integrated \ha \ SFRs of Galaxy Clusters \label{cljlitsfrs}}
\tablehead{\colhead{Name}  	&\colhead{z}  	& \colhead{$\sigma$} & \multicolumn{2}{c}{$R_{200}$} & \multicolumn{2}{c}{Survey Radius} & \colhead{SFR Cor} & \colhead{$\Sigma$~SFR}& \colhead{$\Sigma$~SFR/M$_{cl}$} & \colhead{Tech} & \colhead{Ref.} \\ \colhead{(1)} & (2) & (3) & (4) & (5) & (6) & (7) & (8) & (9) & (10) & (11) &(12)} 
\startdata 
\ca & 0.704 &  418 $\pm$  50 &  1.63 &  0.49 &  1.19 &  0.73 &  1.00 &  30.9 $\pm$  9.28 & 52.18 $\pm$ 15.66 & I & 1 \\ 
\cb & 0.748 &  504 $\pm$  89 &  1.86 &  0.57 &  1.31 &  0.70 &  1.06 &  13.1 $\pm$  3.94 & 12.97 $\pm$  3.89 & I & 1 \\ 
\cc & 0.794 & 1018 $\pm$  75 &  3.59 &  1.13 &  1.15 &  0.32 &  1.16 & 156.9 $\pm$ 47.08 & 19.33 $\pm$  5.80 & I & 1 \\ 
\cj & 0.845 &  415 $\pm$  82 &  1.39 &  0.45 &  1.29 &  0.93 &  1.00 &  38.2 $\pm$  5.00 & 71.60 $\pm$  9.37 & I & 2 \\ 
Abell~2390 & 0.228 & 1023 $\pm$ 102 & 10.29 &  1.58 &  8.00 &  0.78 &  1.70 &  79.9 $\pm$  8.50 &  6.97 $\pm$  0.74 & I & 3 \\ 
AC~114 & 0.320 & 1390 $\pm$ 139 & 10.42 &  2.04 &  4.35 &  0.42 &  2.83 &  21.8 $\pm$ 14.14 &  0.80 $\pm$  0.52 & S & 4 \\ 
Abell~1689 & 0.183 & 1274 $\pm$ 127 & 15.59 &  2.01 &  4.35 &  0.28 &  3.78 &  40.5 $\pm$ 18.92 &  1.78 $\pm$  0.83 & S & 5 \\ 
CL0024.0$+$1652 & 0.395 &  561 $\pm$  89 &  3.52 &  0.79 & 15.00 &  4.26 &  1.00 & 124.0 $\pm$ 17.00 & 72.01 $\pm$  9.87 & I & 6 \\
\enddata\tablecomments{Columns: (1) Cluster name. (2) Redshift. (3) Velocity dispersion in $\rm km\ s^{-1}$. Velocity dispersions for AC~114 and Abell~1689 are calculated from $L_X$ using best-fit $L_X - \sigma$ relation of {Mahdavi} \& {Geller} (2001) because measured dispersions are inflated by substructure. (4) $R_{200}$ in arcmin. (5) $R_{200}$ in \h \ Mpc. (6) Survey radius in arcmin. (7) Survey radius in units of $R_{200}$. (8) Total correction applied to integrated SFR to account for incomplete sampling within $0.5 \times R_{200}$, aperture corrections, and different EW limits.  (9) Integrated SFR in \smy. (10) Integrated SFR per cluster mass, in units of $h_{100}^{-3}\ \rm M_\odot \ yr^{-1}\ / \ 10^{14} \  M_\odot$. (11) Observing Technique: I = narrowband imaging, S = spectroscopic survey. (12) References.} 
\end{deluxetable} 


\begin{references}

\reference {}{Balogh}, M., {Eke}, V., {Miller}, C., {Lewis}, I., {Bower}, R., {Couch}, W.,  {Nichol}, R., {Bland-Hawthorn}, J., {et al.} 2004, \mnras, 348, 1355

\reference {}{Balogh}, M.~L., {Couch}, W.~J., {Smail}, I., {Bower}, R.~G., \& {Glazebrook},  K. 2002, \mnras, 335, 10

\reference {}{Balogh}, M.~L. \& {Morris}, S.~L. 2000, \mnras, 318, 703

\reference {}{Balogh}, M.~L., {Morris}, S.~L., {Yee}, H.~K.~C., {Carlberg}, R.~G., \&  {Ellingson}, E. 1997, \apjl, 488, L75+

\reference {}{Balogh}, M.~L., {Schade}, D., {Morris}, S.~L., {Yee}, H.~K.~C., {Carlberg},  R.~G., \& {Ellingson}, E. 1998, \apjl, 504, L75+

\reference {}{Barbaro}, G. \& {Poggianti}, B.~M. 1997, \aap, 324, 490

\reference {}{Barton}, E.~J., {Geller}, M.~J., \& {Kenyon}, S.~J. 2000, \apj, 530, 660

\reference {}{Bekki}, K. \& {Couch}, W.~J. 2003, \apjl, 596, L13

\reference {}{Bertin}, E. \& {Arnouts}, S. 1996, \aaps, 117, 393

\reference {}{Brinchmann}, J., {Charlot}, S., {White}, S.~D.~M., {Tremonti}, C.,  {Kauffmann}, G., {Heckman}, T., \& {Brinkmann}, J. 2004, \mnras, 351, 1151

\reference {}{Campins}, H., {Rieke}, G.~H., \& {Lebofsky}, M.~J. 1985, \aj, 90, 896

\reference {}{Conselice}, C.~J., {Bershady}, M.~A., \& {Gallagher}, J.~S. 2000, \aap, 354,  L21

\reference {}{Couch}, W.~J., {Balogh}, M.~L., {Bower}, R.~G., {Smail}, I., {Glazebrook}, K.,  \& {Taylor}, M. 2001, \apj, 549, 820

\reference {}{Dressler}, A., {Smail}, I., {Poggianti}, B.~M., {Butcher}, H., {Couch}, W.~J.,  {Ellis}, R.~S., \& {Oemler}, A.~J. 1999, \apjs, 122, 51

\reference {}{Ellingson}, E., {Lin}, H., {Yee}, H.~K.~C., \& {Carlberg}, R.~G. 2001, \apj,  547, 609

\reference {}{Finn}, R.~A., {Balogh}, M.~L., {Miller}, C., {Nichols}, R., \& {Zaritsky}, D.  2005, in prep.

\reference {}{Finn}, R.~A., {Zaritsky}, D., \& {McCarthy}, D.~W. 2004, \apj, 604, 141

\reference {}{G{\' o}mez}, P.~L., {Nichol}, R.~C., {Miller}, C.~J., {Balogh}, M.~L., {Goto},  T., {Zabludoff}, A.~I., {Romer}, A.~K., {Bernardi}, M., {et al.} 2003, \apj,  584, 210

\reference {}{Glazebrook}, K., {Blake}, C., {Economou}, F., {Lilly}, S., \& {Colless}, M.  1999, \mnras, 306, 843

\reference {}{Gonzalez}, A.~H., {Zaritsky}, D., {Dalcanton}, J.~J., \& {Nelson}, A. 2001,  \apjs, 137, 117

\reference {}{Gunn}, J.~E. \& {Gott}, J.~R.~I. 1972, \apj, 176, 1

\reference {}{Halliday}, C., {Milvang-Jensen}, B., {Poirier}, S., {Poggianti}, B.~M.,  {Jablonka}, P., {Arag{\' o}n-Salamanca}, A., {Saglia}, R.~P., {De Lucia}, G., {et al.} 2004, \aap, 427, 397

\reference {}{Homeier}, N., {Demarco}, R., {Postman}, M., {Blakeslee}, J.~P., {Bouwens},  R.~J., {Bradley}, L.~D.~{Ford}, H.~C., {Goto}, R., {Gronwall}, {et al.} 2005, \apj, 621, 651 

\reference {}{Homeier}, N. \& {Gallagher}, J.~S. 2002, \apss, 281, 417

\reference {}{Jansen}, R.~A., {Franx}, M., \& {Fabricant}, D. 2001, \apj, 551, 825

\reference {}{Kauffmann}, G., {Heckman}, T.~M., {White}, S.~D.~M., {Charlot}, S.,  {Tremonti}, C., {Brinchmann}, J., {Bruzual}, G., {Peng}, E.~W., {et al.} 2003,  \mnras, 341, 33

\reference {}{Kauffmann}, G., {White}, S.~D.~M., {Heckman}, T.~M., {M{\' e}nard}, B.,  {Brinchmann}, J., {Charlot}, S., {Tremonti}, C., \& {Brinkmann}, J. 2004,  \mnras, 353, 713

\reference {}{Kennicutt}, R.~C. 1983, \apj, 272, 54

\reference {}---. 1992a, \apjs, 79, 255

\reference {}---. 1992b, \apj, 388, 310

\reference {}---. 1998, \araa, 36, 189

\reference {}{Kennicutt}, R.~C., {Tamblyn}, P., \& {Congdon}, C.~E. 1994, \apj, 435, 22

\reference {}{Keres}, D., {Katz}, N., {Weinberg}, D.~H., \& {Dav{\'e}}, R. 2004, \mnras,  submitted, astroph/0407095

\reference {}{Kodama}, T., {Balogh}, M.~L., {Smail}, I., {Bower}, R.~G., \& {Nakata}, F.  2004, \mnras, in press

\reference {}{Kodama}, T. \& {Bower}, R.~G. 2001, \mnras, 321, 18

\reference {}{Labb{\' e}}, I., {Franx}, M., {Rudnick}, G., {Schreiber}, N.~M.~F., {Rix}, H.,  {Moorwood}, A., {van Dokkum}, P.~G., {van der Werf}, P., {et al.} 2003, \aj, 125, 1107

\reference {}{Larson}, R.~B., {Tinsley}, B.~M., \& {Caldwell}, C.~N. 1980, \apj, 237, 692

\reference {}{Le F{\` e}vre}, O., {Abraham}, R., {Lilly}, S.~J., {Ellis}, R.~S.,  {Brinchmann}, J., {Schade}, D., {Tresse}, L., {Colless}, M., {et al.} 2000, \mnras, 311, 565

\reference {}{Lewis}, I., {Balogh}, M., {De Propris}, R., {Couch}, W., {Bower}, R., {Offer},  A., {Bland-Hawthorn}, J., {Baldry}, I.~K., {et al.} 2002,  \mnras, 334, 673

\reference {}{Mahdavi}, A. \& {Geller}, M.~J. 2001, \apjl, 554, L129

\reference {}{Mannucci}, F., {Basile}, F., {Poggianti}, B.~M., {Cimatti}, A., {Daddi}, E.,  {Pozzetti}, L., \& {Vanzi}, L. 2001, \mnras, 326, 745

\reference {}{McCarthy}, D.~W., {Ge}, J., {Hinz}, J.~L., {Finn}, R.~A., \& {de Jong}, R.~S.  2001, \pasp, 113, 353

\reference {}{Nelson}, A.~E., {Gonzalez}, A.~H., {Zaritsky}, D., \& {Dalcanton}, J.~J. 2001,  \apj, 563, 629

\reference {}{Pell\'o}, R., {Rudnick}, G., {Simard}, L., {White}, S.~D., {Halliday}, C.,  {Milvang-Jensen}, B., {Poirier}, S., {Poggianti}, B.~M., {et al.} 2005, in~prep

\reference {}{Persson}, S.~E., {Murphy}, D.~C., {Krzeminski}, W., {Roth}, M., \& {Rieke},  M.~J. 1998, \aj, 116, 2475

\reference {}{Poggianti}, B.~M., {Smail}, I., {Dressler}, A., {Couch}, W.~J., {Barger},  A.~J., {Butcher}, H., {Ellis}, R.~S., \& {Oemler}, A.~J. 1999, \apj, 518, 576

\reference {}{Poggianti}, B.~M. \& {Wu}, H. 2000, \apj, 529, 157

\reference {}{Postman}, M., {Lubin}, L.~M., \& {Oke}, J.~B. 1998, \aj, 116, 560

\reference {}---. 2001, \aj, 122, 1125

\reference {}{Postman}, M.,  2005, 

\reference {}{Rudnick}, G., {Rix}, H., {Franx}, M., {Labb{\' e}}, I., {Blanton}, M.,  {Daddi}, E., {F{\" o}rster Schreiber}, N.~M., {Moorwood}, A., {et al.} 2003, \apj, 599, 847

\reference {}{Smail}, I., {Dressler}, A., {Couch}, W.~J., {Ellis}, R.~S., {Oemler}, A.~J.,  {Butcher}, H., \& {Sharples}, R.~M. 1997, \apjs, 110, 213

\reference {}{Smith}, et al. 2004, 

\reference {}{Tresse}, L., {Maddox}, S., {Loveday}, J., \& {Singleton}, C. 1999, \mnras,  310, 262

\reference {}{Tresse}, L., {Maddox}, S.~J., {Le F{\` e}vre}, O., \& {Cuby}, J.-G. 2002,  \mnras, 337, 369

\reference {}{van Dokkum}, P.~G., {Franx}, M., {Fabricant}, D., {Kelson}, D.~D., \&  {Illingworth}, G.~D. 1999, \apjl, 520, L95

\reference {}{White}, S.~D., {Clowe}, D., {Simard}, L., {Rudnick}, G., {De~Lucia}, G.,  {Aragon-Salamanca}, A., {Bender}, R., {Best}, P., {et al.} 2004, A\&A, submitted

\reference {}{Zabludoff}, A.~I. \& {Mulchaey}, J.~S. 1998, \apj, 496, 39

\end{references}
\end{document}